\documentclass[preprint2]{aastex}

\usepackage{color}
\usepackage{epsfig}
\usepackage[fleqn]{amsmath}
\usepackage{multirow}
\usepackage{multicol}
\usepackage{mathptmx}
\usepackage{graphicx}
\usepackage{bm}
\usepackage{mathrsfs}

\shorttitle{A Note on the Construction of Explicit
Symplectic Integrators}
\shortauthors{Zhou et al.}
\begin{document}
\title{\large A Note on the Construction of Explicit Symplectic Integrators for Schwarzschild
Spacetimes}
\author{Naying Zhou$^{1,2}$, Hongxing Zhang$^{1,2}$, Wenfang Liu$^{1}$, Xin Wu$^{1,2,3, \dag}$}
\affil{ 1. School of Mathematics, Physics and Statistics, Shanghai
University of Engineering Science, Shanghai 201620, China \\
2. Center of Application and Research of Computational Physics,
Shanghai University of Engineering Science, Shanghai 201620, China \\
3. Guangxi Key Laboratory for Relativistic Astrophysics, Guangxi
University, Nanning 530004, China} \email{$\dag$ Corresponding
Author Email: wuxin$\_$1134@sina.com (X. W.);
M130120101@sues.edu.cn (N. Z.); M130120111@sues.edu.cn (H. Z.);
21200007@sues.edu.cn (W. L.)}

\begin{abstract}

In recent publications, the construction of explicit symplectic
integrators for Schwarzschild and Kerr type spacetimes is based on
splitting and composition methods for numerical integrations of
Hamiltonians or time-transformed Hamiltonians associated with
these spacetimes. Such splittings are not unique but have various
choices. A Hamiltonian describing the motion of charged particles
around the Schwarzschild  black hole with an external magnetic
field can be separated into three, four and five explicitly
integrable parts. It is shown through numerical tests of regular
and chaotic orbits that the three-part splitting method is the
best one of the three Hamiltonian splitting methods in accuracy.
In the three-part splitting, optimized fourth-order partitioned
Runge-Kutta and Runge-Kutta-Nystr\"{o}m explicit symplectic
integrators exhibit the best accuracies. In fact, they are several
orders of magnitude better than the fourth-order Yoshida
algorithms for appropriate time steps. The former algorithms need
small additional computational cost compared with the latter ones.
Optimized sixth-order partitioned Runge-Kutta and
Runge-Kutta-Nystr\"{o}m explicit symplectic integrators have no
dramatic advantages over the optimized fourth-order ones in
accuracies during long-term integrations due to roundoff errors.
The idea finding the integrators with the best performance is also
suitable for Hamiltonians or time-transformed Hamiltonians of
other curved spacetimes including the Kerr type spacetimes. When
the numbers of explicitly integrable splitting sub-Hamiltonians
are as small as possible, such splitting Hamiltonian methods would
bring better accuracies. In this case, the optimized fourth-order
partitioned Runge-Kutta and Runge-Kutta-Nystr\"{o}m methods are
worth recommending.

\end{abstract}

\emph{Unified Astronomy Thesaurus concepts}: Black hole physics
(159); Computational methods (1965); Computational astronomy
(293); Celestial mechanics (211)

\section{Introduction}

Four basic black hole spacetimes consisting of Schwarzschild,
Reissner-Nordstr\"{o}m, Kerr and Kerr-Newman metrics are
integrable. Although their analytical solutions exist from a
theoretical point of view, they cannot be expressed in terms of
elementary functions of time and are only formal solutions
described by elliptic integrals. Numerical techniques are
necessary to  study geodesic orbits of particles in these
spacetimes. When the spacetimes or their corresponding modified
theories of gravity (Deng $\&$ Xie 2016; Deng 2020; Gao $\&$ Deng
2021) contain electromagnetic fields or are immersed in external
electromagnetic fields  acting as perturbations, they become
nonintegrable in many situations. The onset of chaos is even
allowed in the nonintegrable systems (Karas $\&$ Vokroulflick\'{y}
1992; Takahashi $\&$ Koyama 2009; Kop\'{a}\v{c}ek $\&$ Karas 2014;
Stuchl\'{i}k $\&$ Kolo\v{s} 2016; Tursunov et al. 2016;
Kop\'{a}\v{c}ek $\&$ Karas 2018; Panis et al. 2019; Stuchl\'{i}k
et al. 2020). Numerical techniques are particularly important to
solve the nonintegrable problems. Because the spacetimes can
exactly correspond to Hamiltonian systems, their most appropriate
solvers should maintain the symplectic nature of Hamiltonian
dynamics. They are symplectic schemes (Wisdom 1982; Ruth 1983;
Feng 1986; Suzuki 1991). They nearly preserve the energy of a
conservative mechanical system when truncation errors act as a
main error source (see e.g., Hairer et al. 2006). They can also
provide reliable results to the integrated trajectories and to the
detection of the chaotical behavior for appropriate choices of
time steps.

Symplectic methods are divided into explicit symplectic algorithms
and implicit ones (Yoshida 1993). Sometimes their compositions
(i.e., explicit and implicit symplectic composition methods) are
used in the literature (Liao 1997; Preto $\&$ Saha 2009; Lubich et
al. 2010; Zhong et al. 2010; Mei et al. 2013a, 2013b). The
implicit methods do not need splitting the above-mentioned
Hamiltonians, and thus are always available. The implicit midpoint
rule (Feng 1986; Brown 2006) and Gauss-Runge-Kutta methods
(Kop\'{a}\v{c}ek et al. 2010; Seyrich $\&$ Lukes-Gerakopoulos
2012; Seyrich 2013) are common implicit symplectic schemes. They
should be generally more expensive in computational cost than the
explicit and implicit composition methods at same order. Of
course, the latter is more computationally demanding than the
explicit algorithms. Many explicit symplectic algorithms usually
rely on splitting nonlinear Hamiltonians and composing the flows
of the splitting terms. In fact, they are splitting and
composition methods for the numerical integration of nonlinear
ordinary differential equations. There are a class of explicit
symplectic integrators for solving non-separable nonlinear
Hamiltonians, which are the product of a function of momenta and
another function of coordinates and do not need splittings (Chin
2009). In addition, variational symplectic integrators (Marsden
$\&$ West 2001) can be explicit for some nonlinear Hamiltonians
without the use of splitting. The implicit midpoint rule as one of
the variational symplectic integrators becomes explicit for linear
Hamiltonian problems without any splittings.

In general, an $N$-body Hamiltonian problem in the solar system
has a classical splitting into two explicitly integrable parts
with comparable size, which involve a kinetic energy depending on
momenta and a potential energy depending on position coordinates
(Ruth 1983). It can also be split into two explicitly integrable
terms of different magnitudes, i.e., a primary Kepler part and a
small perturbation part corresponding to the interactions among
planets (Wisdom $\&$ Holman 1991). This Hamiltonian splitting is a
perturbative Hamiltonian splitting method. Explicit symplectic
methods, such as fourth-order symplectic integrations of Forest
$\&$ Ruth (1990) and higher-order symplectic algorithms of Yoshida
(1990), are easily feasible and applicable to the classical
splitting and the perturbative splitting. Optimized higher-order
partitioned Runge-Kutta (PRK) and Runge-Kutta-Nystr\"{o}m (RKN)
explicit symplectic integrators (Blanes $\&$ Moan 2002) are
applied for the classical splitting method. However, the
pseudo-higher-order symplectic integrators (Chambers $\&$ Murison
2000; Laskar $\&$ Robutel 2001; Blanes et al. 2013) are
specifically designed for the perturbative splitting. A splitting
of the $N$-body Hamiltonian into three explicitly integrable terms
corresponding to various magnitudes was considered by Duncan et
al. (1998). The pseudo-higher-order symplectic schemes still work
well for such a splitting (Wu et al. 2003). Recently, Chen et al.
(2021) split an N-rigid-body Hamiltonian problem into three and
four integrable terms with various magnitudes and different
timescales. Based on coefficient combinations optimized, the
integration accuracy and efficiency are typically improved in the
splittings.

The Hamiltonians corresponding to the above-mentioned black-hole
spacetimes in general relativity are inseparable to the phase
space variables. In spite of this, the above-mentioned explicit
symplectic integration algorithms or explicit and implicit
combined symplectic methods are always available. When each of the
Hamiltonians is split into two terms, not both terms are
integrable or have analytical solutions as explicit functions of
time. The construction and application of explicit symplectic
integrators is difficult for the splitting method. Doubling the
phase space variables in any inseparable Hamiltonian problem,
Pihajoki (2015) introduced a new Hamiltonian on an extended phase
space with two splitting parts equal to the original inseparable
Hamiltonian system. Here, one part depends on the original
coordinates and the new momenta, and the other part is a function
of the original momenta and the new coordinates. The extended
phase-space Hamiltonian is amenable for integration with a
standard explicit symplectic leapfrog symplectic method. Because
both solutions from the leapfrog integrating the two splitting
parts are coupled through the derivatives, they should diverge
with time. Mixing maps acting as feedback between the two
solutions are necessarily included in the leapfrog so as to solve
this problem. If the mixing maps are nonsymplectic, then the
resulting algorithm no longer symplectic on the extended phase
space. Even if the mixing maps are symplectic, the extended phase
space leapfrog is not symplectic when a solution in the extended
phase space is projected back to that in the original phase space
in any case. The best choice of the mixing maps and projection map
is that the mixing maps take permutations of momenta, and the
projection map takes the original coordinates and the new momenta
in the extended phase space as the solution in the original phase
space. In this way, the explicit extended phase space algorithm
has good long term stability and error behavior although it does
not retain the symplecticity, as was numerically confirmed by
Pihajoki. Thus, it is a symmetrically symplectic-like method. Liu
et al. (2016) pointed out the preference of  sequent permutations
of coordinates and momenta over the permutations of momenta, and
proposed higher order explicit extended phase space
symplectic-like integrators for inseparable Hamiltonian systems.
Luo et al. (2017) found that the midpoint permutations between the
original coordinates and the extended coordinates and the midpoint
permutations between the original momenta and the extended momenta
are the best mixing maps. The explicit extended phase space
symplectic-like integrators with the midpoint permutations are
well applicable to nonconservative nonseparable systems (Luo $\&$
Wu 2017), logarithmic Hamiltonians (Li $\&$ Wu 2017), relativistic
core-shell spacetimes (Liu et al. 2017), and magnetized
Ernst-Schwarzschild spacetimes (Li $\&$ Wu 2019). Recently, Pan et
al. (2021) considered the construction of semiexplicit  extended
phase space symplectic-like integrators for coherent
post-Newtonian Euler-Lagrange equations. On the other hand, Tao
(2016a) did not adopt any mixing maps and proposed explicit
symplectic methods of any even order for a nonseparable
Hamiltonian in an extended phase space. In his method, an extended
phase space Hamiltonian consists of three parts: two copies of the
original system with mixed-up positions and momenta and an
artificial restraint with a parameter $\omega$ controlling  the
binding of the two copies. There are two problems. One problem is
that, although the integrators based on the idea of Tao are
symplectic in the extended phase space, it is unclear that how the
symplecticity of the extended phase space Hamiltonian is related
to that of the original system (Jayawardana $\&$ Ohsawa 2021).
Another problem is that there is no universal method to find the
optimal control parameter $\omega$. The optimal choice relies on
only a large number of numerical tests (Wu $\&$ Wu 2018).
Combining an extended phase space approach of Pihajoki and a
symmetric projection method, Jayawardana $\&$ Ohsawa (2021) have
more recently constructed a semiexplicit symplectic integrator for
inseparable Hamiltonian systems. The computations of the main time
evolution for two copies of the original system with mixed-up
positions and momenta are explicit. However, the computations of
the symmetric projection that binds potentially diverging copies
of solutions are implicit. The resulting method is symplectic in
the original phase space.

In fact, it is possible to construct explicit symplectic
integrators for the aforementioned curved spacetimes in terms of
splitting and composition. One way is a splitting of the
Hamiltonians corresponding to these curved spacetimes into more
explicitly integrable terms of comparable sizes or different
magnitudes. When the Hamiltonian of Schwarzschild spacetime is
separated into four integrable splitting parts with analytical
solutions as explicit functions of proper time, explicit
symplectic methods are easily designed (Wang et al. 2021a).  The
explicit symplectic methods are also suitable for a splitting of
the Hamiltonian of Reissner-Nordstr\"{o}m black hole into five
explicitly integrable terms (Wang et al. 2021b), and  a splitting
of the Hamiltonian of Reissner-Nordstr\"{o}m-(anti)-de Sitter
black hole into six explicitly integrable parts (Wang et al.
2021c). These explicit symplectic methods are still effective when
external magnetic fields are included to destroy the integrability
of these spacetimes. Unfortunately, no such a similar splitting
exists in the Hamiltonian of Kerr black hole and then explicit
symplectic methods do not work. Using a time transformation method
introduced in the work of Mikkola (1997), Wu et al. (2021) gave a
time-transformed Hamiltonian to the Hamiltonian of Kerr black
hole. The time-transformed Hamiltonian is separated into five
explicitly integrable terms and allows for the application of
explicit symplectic methods. This idea was extended to study the
chaotic motions of charged particles around the Kerr black holes
and deformed Schwarzschild black holes immersed in external
magnetic fields (Sun  et al. 2021a, 2021b; Zhang et al. 2021). How
to choose time transformation functions is dependent on some
specific spacetimes. How to split the Hamiltonians or
time-transformed Hamiltonians also depends on the specific
spacetimes.

Generally, for a splitting of a certain Hamiltonian into many
explicitly integrable terms of comparable sizes, the higher-order
explicit symplectic methods of Yoshida (1990) are conveniently
applied. The optimized fourth- and sixth-order PRK and RKN
explicit symplectic integrators (Blanes $\&$ Moan 2002) for the
two-part splitting with comparable size can be adjusted as those
that are appropriate for the multi-part splitting of comparable
sizes (Blanes et al. 2008, 2010). An optimized fourth-order PRK
integrator was recently discussed in the splitting method
(McLachlan 2021). Compared with the Yoshida constructions, the
same order PRK or RKN methods contain more additional time
coefficients and more compositions of all sub-Hamiltonian flows.
For instance, a fourth-order Yoshida method is a symmetric
composition of three second-order leapfrogs, and the optimized
fourth-order PRK integrator is that of six pairs of first-order
approximation composing all the sub-Hamiltonian flows and the
adjoint of the first-order integrator. As a result, the optimized
PRK and RKN methods are somewhat more expensive in computations
than the same order Yoshida integrators.

Now, there is a question of whether the splitting methods of the
above-mentioned or non-mentioned Hamiltonians corresponding to
curved spacetimes are unique. If they are not, which of them
perform the best accuracies. What performances do the optimized
PRK and RKN methods have in various splittings? Are the optimized
PRK and RKN methods superior to the same order Yoshida integrators
in accuracies? To answer these questions, we consider a
Hamiltonian describing the motion of charged particles around the
Schwarzschild  black hole with an external magnetic field as an
example. Besides the splitting of four explicitly integrable parts
introduced in the work of Wang et al. (2021a), two splitting
methods of three  and five explicitly integrable parts with
comparable sizes will be given to the Hamiltonian. Then, the
fourth- and sixth-order Yoshida algorithms and the fourth- and
sixth-order optimized PRK and RKN methods are numerically
evaluated in the three splittings. These algorithms combining the
explicit extended phase space symplectic methods of Tao (2016a) or
the explicit extended phase space symplectic-like integrators with
the midpoint permutations of Luo et al. (2017) are numerically
compared. In a word, the fundamental aim of the present paper is
to find the best integrators and splitting method.

The paper is organized as follows. In Section 2, we introduce
three splitting methods to a Hamiltonian system describing the
motion of charged particles around the Schwarzschild  black hole
with an external magnetic field. Yoshida algorithms and optimized
PRK and RKN methods of orders 4 and 6 are used in the three
splittings.  In Section 3, we check the numerical performance of
these algorithms in the three splitting methods. Finally, the main
results are concluded in Section 4. Some explicit extended phase
space symplectic or symplectic-like methods are described in
Appendix.

\section{Splitting Hamiltonian methods and explicit Symplectic integrators}

First, we present a Hamiltonian dynamical system for the
description of charged particles moving around the Schwarzschild
black hole with an external magnetic field. Second, an existing
splitting of the Hamiltonian into four explicitly integrable terms
is introduced, and Yoshida algorithms and optimized PRK and RKN
methods of orders 4 and 6 are applied to the splitting. Third, a
splitting into three explicitly integrable parts is given to the
Hamiltonian and these integrators are considered in such a
splitting. Finally, the mentioned integrators act on a splitting
of the Hamiltonian into five explicitly integrable parts.

\subsection{Hamiltonian formulism for Schwarzschild spacetime with external magnetic field}

The dynamics of a test particle with charge $q$ moving around the
Schwarzschild black hole surrounded by an external magnetic field
is described by the following Hamiltonian (Kolo\v{s} et al. 2015)
\begin{eqnarray}
H =\frac{1}{2}g^{\mu\nu}(p_{\mu}-qA_{\mu})(p_{\nu}-qA_{\nu}).
\end{eqnarray}
In spherical-like  coordinates $(t, r, \theta, \varphi)$, nonzero
components of the Schwarzschild metric $g^{\mu\nu}$ are
\begin{eqnarray}
g^{tt} &=& -(1-\frac{2}{r})^{-1}, ~~~~~~ g^{rr}=1-\frac{2}{r},
\nonumber \\
g^{\theta\theta} &=& \frac{1}{r^2}, ~~~~~~~~~~~~~~~~~~~
g^{\varphi\varphi}=\frac{1}{r^2\sin^2\theta}. \nonumber
\end{eqnarray}
The external uniform magnetic field  in the vicinity of the black
hole has an electromagnetic field potential with only one nonzero
covariant component (Kolo\v{s} et al. 2015; Panis et al. 2019)
\begin{eqnarray}
   A_{\varphi}=\frac{B}{2}r^{2}\sin^{2}\theta,
\end{eqnarray}
where $B$ represents a magnetic field strength. Here, a point on
the presence of magnetic field in the vicinity of the black hole
is illustrated. Observations show that strong magnetic fields
exist in active galactic nuclei (Xu et al. 2011). A regular
magnetic field might arise inside an accretion disk around a black
hole due to the dynamo mechanism in conducting matter (plasma) of
the accretion disk (Tursunov et al. 2013; Abdujabbarov et al.
2014). This magnetic field does not get through the conducting
plasma region and falls in the vicinity of the black hole (Frolov
2012). At large distances, the character of a large-scale magnetic
field in accretion processes can be approximately simplified to a
homogeneous magnetic field in finite element of space. For
simplicity, an asymptotically uniform magnetic field is considered
as the external magnetic field (Wald 1974; Kov\'{a}\v{r} et al.
2014; Stuchl\'{i}k  $\&$ Kolo\v{s} 2016).  $p_{\mu}$ is a
generalized momentum determined by a set of canonical Hamiltonian
equations $\dot{x}^{\mu}=\partial H/\partial p_{\mu}$, and reads
\begin{eqnarray}
  p_{\mu}=g_{\mu\nu}\dot{x}^{\nu}+qA_{\mu}.
\end{eqnarray}
Here, covariant metric components are $g_{tt}=1/g^{tt}$,
$g_{rr}=1/g^{rr}$, $g_{\theta\theta}=1/g^{\theta\theta}$ and
$g_{\varphi\varphi}=1/g^{\varphi\varphi}$. 4-velocity
$\dot{x}^{\nu}$ is a derivative of coordinate $x^{\nu}$ with
respect to proper time $\tau$.

Another set of canonical Hamiltonian equations
$\dot{p}_{\mu}=-\partial H/\partial \dot{x}^{\mu}$ show
$\dot{p}_{t}=\dot{p}_{\varphi}=0$. Namely, two constant
generalized momentum components are
\begin{eqnarray}
p_{t} &=& g_{tt}\dot{t}=-(1-\frac{2}{r})\dot{t}=-E, \\
p_{\varphi} &=&
g_{\varphi\varphi}\dot{\varphi}+qA_{\varphi}=r^{2}\sin^{2}\theta(\dot{\varphi}+\frac{\beta}{2})=L,
\end{eqnarray}
where $\beta=qB$. $E$ is a constant energy of the particle, and
$L$ corresponds to a constant angular momentum of the particle.
Substituting the two constants into Equation (1), we rewrite the
Hamiltonian as
\begin{eqnarray}
H  &=&
\frac{1}{2}(1-\frac{2}{r})p_{r}^{2}-\frac{1}{2}(1-\frac{2}{r})^{-1}E^{2}
+\frac{p_{\theta}^{2}}{2r^{2}}
\nonumber \\
&& +\frac{1}{2r^{2}\sin^{2}\theta}(L-\frac{\beta}{2}
r^{2}\sin^{2}\theta)^{2}.
\end{eqnarray}
Due to the particle's rest mass in the time-like spacetime, a
third constant is always given by
\begin{eqnarray}
H  = -\frac{1}{2}.
\end{eqnarray}
No fourth constant exists when the magnetic field is included in
the Schwarzschild spacetime. Thus, the Hamiltonian (6) is a
nonintegrable system with two degrees of freedom in a
four-dimensional phase space.

A point is illustrated here.  The speed of light $c$ and the
gravitational constant $G$ are measured in terms of geometric
units, $c=G=1$. Equation (6) with Equation (7) is dimensionless.
The dimensionless operations to the related qualities are
implemented through scale transformations to the qualities. That
is, $t \rightarrow tM$, $\tau \rightarrow \tau M$, $ r\rightarrow
rM$, $ B\rightarrow B/M$, $E\rightarrow mE$, $p_r\rightarrow
mp_r$, $L\rightarrow mML$, $p_\theta\rightarrow mMp_\theta$,
$q\rightarrow mq$ and $H\rightarrow m^2H$, where $M$ denotes the
black hole's mass and $m$ stands for the particle's mass.

\subsection{An existing splitting method}

Separations of the variables in the Hamiltonian (6), including the
separation of momenta $p_r$ and $p_\theta$ from coordinates $r$
and $\theta$ or the separation of variables $r$ and $p_r$ from
variables $\theta$ and $p_\theta$, are impossible. In spite of
this fact, the Hamiltonian can still be split into two integrable
parts with analytical solutions; e.g., one part is composed of the
first, second and fourth terms, and another part is the third
term. Unfortunately, not both the parts have \emph{explicit}
analytical solutions. Thus, explicit symplectic integrators are
not applicable to the Hamiltonian splitting. However, they are
available when the Hamiltonian is separated into four parts with
explicit analytical solutions explicitly depending on proper time
$\tau$, as was shown by Wang et al. (2021a). In what follows, we
briefly introduce the idea on the construction of explicit
symplectic methods.

Wang et al. (2021a) suggested splitting the Hamiltonian (6) into
four parts as follows:
\begin{eqnarray}
H &=& H_{1}+H_{2}+H_{3}+H_{4}, \\
H_{1} &=& \frac{1}{2r^{2}\sin^{2}\theta}(L-\frac{\beta }{2}r^{2}\sin^{2}\theta )^{2} \nonumber \\
&& -\frac{1}{2}(1-\frac{2}{r})^{-1}E^{2},\\
H_{2} &=& \frac{1}{2}p_{r}^{2},\\
H_{3}&=& -\frac{1}{r}p_{r}^{2},\\
H_{4} &=& \frac{1}{2r^{2}}p_{\theta}^{2}.
\end{eqnarray}
It is clear that each of the four sub-Hamiltonians $H_{1}$,
$H_{2}$, $H_{3}$ and $H_{4}$ is analytically solvable, and its
analytical solutions are explicit functions of proper time $\tau$.
The exact solvers for the four parts are in sequence labelled as
$\Xi_{h}^{H_{1}}$, $\Xi_{h}^{H_{2}}$, $\Xi_{h}^{H_{3}}$ and
$\Xi_{h}^{H_{4}}$, where $h$ is a proper time step.

\subsubsection{Yoshida's Constructions}

The exact flow of Hamiltonian (6) advancing time $h$,
$\Xi_{h}^{H}$, is approximately expressed as
\begin{eqnarray}
\Xi_{h}^{H} \approx S2A(h) &=& \Xi_{h/2}^{H_{4}}\times
\Xi_{h/2}^{H_{3}}
\times \Xi_{h/2}^{H_{2}} \times \Xi_{h}^{H_{1}}  \nonumber \\
  &&  \times \Xi_{h/2}^{H_{2}}\times \Xi_{h/2}^{H_{3}}\times \Xi_{h/2}^{H_{4}}.
\end{eqnarray}
S2 is a symmetric composition product of these solvable operators
$\Xi_{h}^{H_{1}}$, $\Xi_{h/2}^{H_{2}}$, $\Xi_{h/2}^{H_{3}}$ and
$\Xi_{h/2}^{H_{4}}$. It is a second-order explicit symplectic
solver for the Hamiltonian (6). Symmetric products of  S2A solvers
can produce fourth- and sixth-order explicit symplectic schemes
(Yoshida 1990):
\begin{eqnarray}
   S4A(h)=S2A(c_{1}h)\times S2A(c_{2}h)\times S2A(c_{1}h),\\
   S6A=S4A(d_{1}h)\times S4A(d_{2}h)\times S4A(d_{1}h).
\end{eqnarray}
where $c_{1}=1/(2-2^{1/3})$, $c_{2}=1-2c_{1}$, $d_{1}=1/(2-2^{1/5})$ and
$d_{2}=-2^{1/5}/(2-2^{1/5})$.

These explicit symplectic algorithms proposed by  Wang et al.
(2021a) are specifically designed for the Schwarzschild type
spacetimes without or with perturbations from weak external
sources like magnetic fields. The Hamiltonian splitting (8) is
also suitable for the construction of higher order optimized
explicit symplectic algorithms of  Blanes et al. (2010), who
introduced symmetric compositions using more extra stages.

\subsubsection{Optimized symplectic PRK and
RKN methods}

Consider two first-order approximations to the exact solutions of
the system (8):
\begin{eqnarray}
   \chi_{Ah} &=& \Xi_{h}^{H_{4}}\times
   \Xi_{h}^{H_{3}}
   \times \Xi_{h}^{H_{2}} \times \Xi_{h}^{H_{1}}, \\
   \chi_{Ah}^{*} &=& \Xi_{h}^{H_{1}}
     \times \Xi_{h}^{H_{2}}\times \Xi_{h}^{H_{3}}
   \times \Xi_{h}^{H_{4}}.
\end{eqnarray}
Note that $\chi_{Ah}^{*} = \chi _{A(-h)}^{-1}$ is the adjoint of
$\chi_{Ah}$. Using both maps $\chi_{Ah}$ and $\chi_{Ah}^{*}$,
Blanes et al. (2008, 2010) introduced a symmetric composition
\begin{eqnarray}
   \psi_{As} = \chi _{A \alpha_{2s}h}\times \chi^{*}_{A \alpha_{2s-1}h} \times  \cdots \times
     \chi^{*}_{A \alpha_{2}h} \times \chi _{A \alpha_{1}h},
\end{eqnarray}
where a series of coefficients are $\alpha_{0}=\alpha_{2s+1}=0$,
and
\begin{eqnarray}
   \alpha_{1} &=& a_{1}, \\
   \alpha_{2j+1} &=& a_{1}+\sum_{k=1}^{j}(a_{k+1}-b_{k}),\\
   \alpha_{2j} &=& \sum_{k=1}^{j}(b_{k}-a_{k}).
\end{eqnarray}
In the above equations, coefficients $a_{1}$, $\cdots$, $b_{1}$,
$\cdots$ with  $\sum_{i=1}^{s}a_{i}=\sum _{i=1}^{s}b_{i}$ are
stemmed from those of the symmetric fourth- and sixth-order
symplectic partitioned Runge-Kutta (PRK) and
Runge-Kutta-Nystr\"{o}m (RKN) methods for the two-part splitting,
and are listed in Tables 2 and 3 by  Blanes $\&$ Moan (2002).

When $s=1$, Equation (18) is the second-order algorithm (13):
\begin{eqnarray}
   S2A = \chi _{A h/2}\times \chi^{*}_{A h/2}.
\end{eqnarray}
Such a pair of operator $\chi$ and its adjoint $ \chi^{*}$ can
compose higher-order integrators.

Given $s=6$, Equation (18) corresponds to a fourth-order optimal
explicit symplectic PRK algorithm
\begin{eqnarray}
   PRK_{6}4A=\chi _{A \alpha_{12}h}\times \chi^{*}_{A \alpha_{11}h} \cdots
     \chi^{*}_{A \alpha_{2}h} \times \chi _{A \alpha_{1}h},
\end{eqnarray}
where $\alpha_{1}$, $\cdots$, and $\alpha_{12}$ calculated by us
are given in Table 1. The optimization means that free
coefficients among coefficients $a_{i}$, $b_{i}$  minimize the
truncation errors at the fifth order. The free coefficients arise
because the number of coefficients $a_{i}$, $b_{i}$ is more than
that of the order conditions.  The optimization can drastically
lead to reducing discretization errors at fixed cost, compared
with the non-optimization. McLachlan (2021) confirmed that the
ordering of the separable terms in the algorithm affects the
errors and slightly affects the computational cost. Thus, choosing
the best ordering is important to reduce the errors. Clearly, the
optimized fourth-order PRK integrator contain more additional time
coefficients and more compositions of all sub-Hamiltonian flows
than the fourth-order Yoshida method. In fact, the former is a
symmetric composition of six pairs of operator $\chi$ and its
adjoint $\chi^{*}$, and the latter is that of three second-order
methods S2A.

For $s=10$ in Equation (18), a sixth-order optimal explicit
symplectic PRK method is
\begin{eqnarray}
  PRK_{10}6A = \chi _{A \alpha_{20}h}\times \chi^{*}_{A \alpha_{19}h} \cdots
     \chi^{*}_{A \alpha_{2}h} \times \chi _{A \alpha_{1}h},
\end{eqnarray}
where the values of $\alpha_{1}$-$\alpha_{20}$ are listed in Table
1. This integrator is a symmetric composition of ten pairs of
operator $\chi$ and its adjoint $ \chi^{*}$.

On the other hand, Equation (18) can also yield RKN methods.
Taking $s=6$, we have a fourth-order optimal explicit symplectic
RKN method
\begin{eqnarray}
  RKN_{6}4A=\chi _{A \alpha_{12}h}\times \chi^{*}_{A \alpha_{11}h} \cdots
     \chi^{*}_{A \alpha_{2}h} \times \chi _{A \alpha_{1}h}.
\end{eqnarray}
Given $s=11$, a sixth-order optimal explicit symplectic RKN method
reads
\begin{eqnarray}
  RKN_{11}6A = \chi _{A \alpha_{22}h}\times \chi^{*}_{A \alpha_{21}h} \cdots
     \chi^{*}_{A \alpha_{2}h} \times \chi _{A \alpha_{1}h}.
\end{eqnarray}
For $s=14$, another sixth-order optimal explicit symplectic RKN
method is
\begin{eqnarray}
  RKN_{14}6A = \chi _{A \alpha_{28}h}\times \chi^{*}_{A \alpha_{27}h} \cdots
     \chi^{*}_{A \alpha_{2}h} \times \chi _{A \alpha_{1}h}.
\end{eqnarray}
We use Equations (19)-(21) to calculate the coefficients of the
three algorithms, which are listed in Table 1.

The PRK and RKN methods for the Hamiltonian splitting (8) need
more compositions of  operators $\Xi_{h}^{H_{1}}$,
$\Xi_{h}^{H_{2}}$, $\Xi_{h}^{H_{3}}$ and $\Xi_{h}^{H_{4}}$ than
the same order Yoshida's constructions. Such a splitting
Hamiltonian method is not unique. There are other splitting
Hamiltonian methods to construct explicit symplectic schemes.

\subsection{Other Hamiltonian splitting methods}

We focus on the application of the aforementioned integrators to
two splittings of the Hamiltonian into three and five explicitly
integrable terms.

\subsubsection{Splitting three parts}

The Hamiltonian (6) can be split into three parts
\begin{eqnarray}
   H=H_{1}+\mathcal{H}_{2}+H_{3},
\end{eqnarray}
where $\mathcal{H}_{2}$ is the sum of $H_2$ and $H_4$:
\begin{eqnarray}
      \mathcal{H}_{2}=\frac{1}{2}p_{r}^{2}+\frac{1}{2r^{2}}p_{\theta}^{2}.
\end{eqnarray}

The canonical  equations of sub-Hamiltonian $ \mathcal{H}_{2}$ are
written as
\begin{eqnarray}
   \frac{dr}{d\tau} &=& \frac{\partial \mathcal{H}_{2}}{\partial
   p_{r}}=p_{r}, ~~ \frac{dp_{r}}{d\tau}=-\frac{\partial \mathcal{H}_{2}}{\partial
   r}=\frac{p^2_{\theta}}{r^{3}}; \\
   \frac{d\theta}{d\tau} &=& \frac{\partial \mathcal{H}_{2}}{\partial
   p_{\theta}}=\frac{p_{\theta}}{r^{2}}, ~~
   \frac{dp_{\theta}}{d\tau}=-\frac{\partial \mathcal{H}_{2}}{\partial
   \theta}=0.
   \end{eqnarray}
They are exactly, analytically solved. Advancing time $h$ from
solutions $(r_{n-1},\theta_{n-1}, p_{rn-1}, p_{\theta n-1})$ at
proper time $\tau_{n-1}$, the analytical solutions at proper time
$\tau_{n}=\tau_{n-1}+h$ are expressed as
\begin{eqnarray}
r_{n} &=& \frac{1}{e_1\cos\theta_{n-1}+e_2\sin\theta_{n-1}}, \\
\theta_{n} &=& f_1+\arctan[(e^{2}_1+e^{2}_2)h p_{\theta _{n-1}}+f_{2}],\\
p_{rn} &=& p_{\theta _{n-1}}(e_1\sin\theta_{n-1}-e_2\cos\theta_{n-1});\\
e_1 &=&
\frac{\cos\theta_{n-1}}{r_{n-1}}+\frac{p_{r_{n-1}}\sin\theta_{n-1}}{p_{\theta
_{n-1}}}, \nonumber \\
e_2 &=&
\frac{\sin\theta_{n-1}}{r_{n-1}}+\frac{p_{r_{n-1}}\cos\theta_{n-1}}{p_{\theta
_{n-1}}},\nonumber \\
f_1 &=&\arctan2(e_2,e_1), \nonumber \\
f_{2} &=& \tan(\theta_{n-1}-f_1).\nonumber
\end{eqnarray}

Equations (13)-(15) are rewritten as
\begin{eqnarray}
S2B(h) = \Xi_{h/2}^{H_{3}} \times \Xi_{h/2}^{\mathcal{H}_{2}}
\times \Xi_{h}^{H_{1}}  \times \Xi_{h/2}^{\mathcal{H}_{2}}\times
\Xi_{h/2}^{H_{3}},
\end{eqnarray}
\begin{eqnarray}
   S4B(h) = S2B(c_{1}h)\times S2B(c_{2}h)\times S2B(c_{1}h),\\
   S6B=S4B(d_{1}h)\times S4B(d_{2}h)\times S4B(d_{1}h).
\end{eqnarray}
Their constructions are based on the Hamiltonian three-part
splitting (28).

Let us define two first-order maps
\begin{eqnarray}
   \chi_{Bh} &=& \Xi^{H_{3}}_{h}
   \times \Xi^{\mathcal{H}_{2}}_{h} \times \Xi^{H_{1}}_{h}, \\
   \chi_{Bh}^{*} &=& \Xi^{H_{1}}_{h}
     \times \Xi^{\mathcal{H}_{2}}_{h}\times \Xi^{H_{3}}_{h}.
\end{eqnarray}
Through $\chi_{Ah}\rightarrow\chi_{Bh}$ and
$\chi^{*}_{Ah}\rightarrow\chi^{*}_{Bh}$, algorithms $PRK_{6}4A$,
$PRK_{10}6A$, $RKN_{6}4A$, $RKN_{11}6A$ and $RKN_{14}6A$ become
methods $PRK_{6}4B$, $PRK_{10}6B$, $RKN_{6}4B$, $RKN_{11}6B$ and
$RKN_{14}6B$, respectively.

\subsubsection{Splitting five parts}

Now, we give five separable parts to the Hamiltonian (6) as
follows:
\begin{eqnarray}
   H=H_{1}+\mathbb{H}_{2}+H_{3}+H_{4}+H_{5},
\end{eqnarray}
where two new sub-Hamiltonians are
\begin{eqnarray}
\mathbb{H}_{2} &=& \frac{1}{2}(1+ar)p_{r}^{2}, \\
H_{5} &=& -\frac{ar}{2}p_{r}^{2}.
\end{eqnarray}
Here, $a$ is a free parameter. Various choices of $a$ correspond
to different five part Hamiltonian decompositions. In other words,
there are an infinite number of methods for the Hamiltonian split
into five parts.

Sub-Hamiltonian $\mathbb{H}_{2}$ corresponds to evolution
equations
\begin{eqnarray}
   \frac{dr}{d\tau} &=& (1+ar)p_{r}, \\
   \frac{dp_{r}}{d\tau} &=& -\frac{a}{2}p_{r}^{2}.
   \end{eqnarray}
The two equations have the analytical solutions
\begin{eqnarray}
   r_{n} &=& \frac{1}{a}[(-\frac{h}{2} ap_{r_{n-1}}(1+ar_{n-1})^{1/2} \nonumber \\
   && +(1+ar_{n-1})^{1/2})^{2}-1],\\
   p_{rn} &=& (-\frac{ah}{2}+\frac{1}{p_{r_{n-1}}})^{-1}.
\end{eqnarray}
For sub-Hamiltonian $H_{5}$, the equations of motion are
\begin{eqnarray}
   \frac{dr}{d\tau} &=& -arp_{r},\\
   \frac{dp_{r}}{d\tau} &=& \frac{a}{2}p_{r}^{2}.
\end{eqnarray}
Their analytical solutions read
\begin{eqnarray}
  r_{n} &=& \frac{1}{a}[-\frac{ahp_{r_{n-1}} r^{1/2}_{n-1}}{2} +r^{1/2}_{n-1}]^{1/2},\\
  p_{r_{n}} &=& (-\frac{ah}{2}+\frac{1}{p_{r_{n-1}}}  )^{-1}.
\end{eqnarray}
Equations (13)-(15) become
\begin{eqnarray}\nonumber
S2C(h) &=\Xi_{h/2}^{H_{5}}\times\Xi_{h/2}^{H_{4}}
\times\Xi_{h/2}^{H_{3}}\times\Xi_{h/2}^{\mathbb{H}_{2}}\times\Xi_{h}^{H_{1}}\\
   &\times\Xi_{h/2}^{\mathbb{H}_{2}}\times\Xi_{h/2}^{H_{3}}\times\Xi_{h/2}^{H_{4}}\times\Xi_{h/2}^{H_{5}},
\end{eqnarray}
\begin{eqnarray}
   S4C(h) = S2C(c_{1}h)\times S2C(c_{2}h)\times S2C(c_{1}h),\\
   S6C=S4C(d_{1}h)\times S4C(d_{2}h)\times S4C(d_{1}h).
\end{eqnarray}

Take two first-order maps
\begin{eqnarray}
   \chi_{Ch} &=& \Xi_{h}^{H_{5}}
   \times \Xi_{h}^{H_{4}} \times \Xi_{h}^{H_{3}}
   \times \Xi_{h}^{\mathbb{H}_{2}} \times \Xi_{h}^{H_{1}}, \\
   \chi^{*}_{Ch} &=& \Xi_{h}^{H_{1}}
     \times \Xi_{h}^{\mathbb{H}_{2}}\times \Xi_{h}^{H_{3}}
     \times \Xi_{h}^{H_{4}}\times \Xi_{h}^{H_{5}}.
\end{eqnarray}
In terms of $\chi_{Ah}\rightarrow\chi_{Ch}$ and
$\chi^{*}_{Ah}\rightarrow\chi^{*}_{Ch}$, algorithms $PRK_{6}4A$,
$PRK_{10}6A$, $RKN_{6}4A$, $RKN_{11}6A$ and $RKN_{14}6A$
correspond to methods $PRK_{6}4C$, $PRK_{10}6C$, $RKN_{6}4C$,
$RKN_{11}6C$ and $RKN_{14}6C$, respectively.

Two points are worth noticing. First, the above-mentioned three,
four and five splitting parts might have comparable sizes
sometimes, or might have various magnitudes and different
timescales. In other words, these splitting parts do not always
have various magnitudes, and should be considered to be comparable
sizes in the whole course of integration. Therefore, the optimized
coefficient combinations in the explicit symplectic integrations
for the N-rigid-body Hamiltonian problem into three and four
integrable terms of various magnitudes and different timescales
(Chen et al. 2021) are not suitable for the present splitting and
composition methods. Second, if the asymptotically uniform
magnetic field in Equation (2) gives place to the general magnetic
fields of Tao (2016b), the present splitting and composition
methods fail to construct the explicit symplectic methods.
However, the Tao's explicit symplectic integrators are still
valid. In fact, the Tao's construction is unlike ours. In the
Tao's method, one of the two Hamiltonian splitting parts has an
analytical solution, whereas another part is not analytically
available and uses Runge-Kutta approximation to calculate
position. In our construction, each of the  Hamiltonian splitting
parts is solved analytically.

\section{Numerical comparisons}

In this section, we mainly check the numerical performance of the
above-mentioned algorithms in the three splitting Hamiltonian
methods. For comparison,  the explicit extended phase space
symplectic-like methods with the midpoint permutations of Luo et
al. (2017) and the explicit extended phase space symplectic
methods without any permutations of Tao (2016a)  are considered.
Their details are introduced in Appendix.

\subsection{Best choice of $a$ in the five splitting parts}

The parameters are $E=0.995$, $L=4.6$, and
$\beta=8.9\times10^{-4}$. The initial conditions are
$\theta=\pi/2$ and $p_{r}=0$; the initial value $p_{\theta}>0$ is
determined by Equation (7). Taking $a=h=1$, we employ the
second-order method S2C to plot Figure 1, which describes two
orbits with the initial separations $r=11$, 72 in Poincar\'{e}
surface of section $\theta=\pi/2$ with $p_{\theta}<0$. The initial
separation $r=11$ corresponds a closed curve, which indicates
regular motion. The motion for the initial separation $r=72$ is
chaotic because the plotted points are randomly distributed in an
area. The orbital regularity or chaoticity for any conservative
Hamiltonian system with two degrees of freedom in four-dimensional
phase space can be seen clearly from the distribution of the
points in the Poincar\'{e} map.

Let us choose the regular orbit with the initial separation $r=11$
as a test orbit to evaluate how a variation of $a$ affects the
numerical performance of S2C. When $a$ ranges from 0 to 10 with an
interval $\Delta a=0.01$, the dependence of Hamiltonian error
$\Delta H=1+2H$ on $a$ is shown in Figure 2(a), where each error
is obtained by S2C after the integration time $\tau=10^7$.
Clearly, $a=1.0260$ corresponds to the minimum error. This result
is also supported in Figure 2(b) on the description of Hamiltonian
errors for $a=1.0260$, 3.9691, 8.6610. Hereafter, $a=1.0260$ is
used in the C type algorithms for the Hamiltonian five-part
splitting (40).

\subsection{Checking numerical performance}

Now, the ordered orbit in Figure 1 is still used as a test orbit.
The second-order, fourth-order and sixth-order symplectic schemes
of Yoshida are applied to the three Hamiltonian splitting methods.
The Hamiltonian errors of these algorithms are described in
Figures 3 (a)-(c). They remain stable and bounded, and do not grow
with time. When truncation errors are larger than roundoff errors,
the boundness of Hamiltonian (or energy) errors in long-term
integrations is the intrinsic property of these symplectic
integrators. Different Hamiltonian splitting methods affect
numerical errors. In accuracies, the fourth-order methods are
better than the second-order ones, but poorer than the sixth-order
ones; at same orders, the B type algorithms are always superior to
the A type methods, which are inferior to the C type integrators.
Namely, the three-part splitting method has the best accuracy,
whereas the five-part splitting method performs the poorest
accuracy for a given integrator. When the ordered orbit in Figure
1 is replaced with the chaotic orbit, the results are still the
absence of error growth for the second- and fourth-order schemes.
The preference of the fourth-order schemes over the second-order
schemes, and the preference of  the B type algorithms over the A
type methods at the second and fourth orders are shown in Figures
3 (d) and (e). However, the errors of the sixth-order method in
the three Hamiltonian splittings are almost the same and begin to
yield a secular drift when the integration time spans $10^5$ in
Figure 3(f). This drift is due to roundoff errors. The truncation
errors are appropriately accurate to the machine double precision
before the integration time $\tau=1000$. As the integration
continues, the roundoff errors slowly increase and dominate the
truncation errors. This leads to the drift in the errors. Thus,
the accuracies of the sixth-order algorithms have no advantage
over those of the fourth-order methods when the integration time
is long enough. Comparisons between Figures 3(a) and 3(d),
comparisons between Figures 3(b) and 3(e), and comparisons between
Figures 3(c) and 3(f) show that each of the integrators for the
same Hamiltonian splitting exhibits better energy accuracy for the
chaotic orbit than for the ordered orbit. This result is due to
the average period of the chaotic orbit larger than that of the
regular orbit. Although the chaotic orbit lacks periodicity, its
average period is admissible. Given a time step, a larger average
orbital period should bring better accuracy. On the contrary, the
accuracy of solutions becomes poorer for the chaotic case than for
the regular case. This is because sensitive dependence of the
solutions  on the initial conditions for the chaotic case must
give rise to the rapid accumulation of  errors of the solutions.

What about the numerical performance of the PRK and RKN
integrators for the three  splitting methods? The ordered orbit in
Figure 1 is still chosen as a test orbit. Figures 4 (a) and (b)
still support the results described in Figures 3 (a)-(e). That is,
the B type algorithms have the best accuracies in the fourth-order
optimized methods $PRK_64$ and $RKN_64$, but the C type algorithms
yield the poorest accuracies. The B type sixth-order methods in
Figures 4 (c) and (d) are better than the A ones within the
integration time $\tau=10^6$. As the integration time lasts long
enough, $PRK_{10}6B$ is inferior to $PRK_{10}6A$, and $RKN_{11}6B$
is inferior to $RKN_{11}6A$ due to the fast growth of roundoff
errors. In Figure 4(e), the energy error for $RKN_{14}6B$ is
smaller than that for $RKN_{14}6A$, and both errors  have no
secular drifts. However, the energy error for $RKN_{14}6B$ has one
or two orders of magnitude larger than for $RKN_{11}6B$ in Figure
4(d) and $PRK_{10}6B$ in Figure 4(c). Thus, the influence of the
roundoff errors on the global errors is smaller for $RKN_{14}6B$
than for $RKN_{11}6B$ and $PRK_{10}6B$. Table 2 lists CPU times of
these algorithms solving the regular orbit in Figures 3 and 4.
The higher the order of an algorithm is, the more CPU time the
algorithm takes. Although CPU time  with 3 minutes 23 seconds for
$RKN_{14}6C$ is more than CPU time with 32 seconds for S2A,
additional CPU time with 2 minutes 51 seconds is still acceptable.
In particular, $PRK_64B$ and $RKN_64B$ have appropriately same CPU
times, and need small additional computational cost compared with
S4B.

The main results of the A and B type algorithms in Figures 3 and 4
are included in Figures 5 (a) and (c). The fourth-order methods
from high accuracy to low accuracy in Figure 5(a) are $PRK_64B$
$\succ$ $RKN_64B$ $\succ$ $PRK_64A$ $\succ$ $RKN_64A$ $\succ$ S4B
$\succ$ S4A. Note that $PRK_64B$ is slightly better than $RKN_64B$
in accuracy. In particular, the errors of $PRK_64B$ and $RKN_64B$
are four orders of magnitude smaller than that of S4A, and three
orders of magnitude smaller than that of S4B. For comparison, the
errors of four methods TS4, LS4, TPRK$_{6}$4 and LPRK$_{6}$4  are
plotted in Figure 5(b). TS4 is the Yoshida's fourth-order
construction combining the extended phase space symplectic method
of Tao (2016a). LS4 is the Yoshida's fourth-order construction
combining the extended phase space symplectic-like method of Luo
et al. (2017). TPRK$_{6}$4 is the fourth-order PRK$_{6}$4 method
combining the Tao's method. LPRK$_{6}$4 is the fourth-order
PRK$_{6}$4 method combining the method of Luo et al. Their details
are given in Appendix. It can be seen clearly from Figures 5 (a)
and (b) that TS4 is almost the same as S4A in accuracy. The
algorithms from low accuracy to high accuracy are TS4 $\prec$ LS4
$\prec$ S4B $\prec$ TPRK$_{6}$4 $\prec$ LPRK$_{6}$4 $\prec$
RKN$_{6}$4B $\prec$ PRK$_{6}$4B. The superiority  of  the extended
phase space symplectic-like methods with the midpoint permutations
to the same type extended phase space symplectic methods without
the use of any permutations (e.g., LS4 $\succ$ TS4) in accuracy is
consistent with that of Wu $\&$ Wu (2018). As far as the CPU times
in Table 2 are concerned, the computational efficiency of LS4 is
slightly superior to that of TS4. The efficiency of TS4 is close
to those of S4A and S4B. The cost of TPRK$_{6}$4 is slightly
larger than those of RKN$_{6}$4B and PRK$_{6}$4B, whereas that of
LPRK$_{6}$4 is slightly smaller. When the integration time is less
than $10^5$, the sixth-order methods from good accuracy to poor
accuracy in Figure 5(c) are $PRK_{10}6B$ $\succ$ $RKN_{11}6B$
$\succ$ $PRK_{10}6A$ $\succ$ $RKN_{11}6A$ $\succ$ $RKN_{14}6B$
$\succ$ S6B $\succ$ $RKN_{14}6A$ $\succ$ S6A. As the integrations
continue, the four methods $PRK_{10}6B$, $RKN_{11}6B$,
$PRK_{10}6A$ and $RKN_{11}6A$ show secular drifts in the errors.
The related errors of these algorithms are clearly listed in Table
3. The fourth-order methods $PRK_64B$ and $RKN_64B$ are four
orders of magnitude better in accuracies than the fourth-order
Yoshida method S4A or the fourth-order Tao extended phase space
method TS4. The sixth-order methods such as $RKN_{11}6B$ have no
advantages over the fourth-order methods $PRK_64B$ and $RKN_64B$.

Let us investigate the accuracies of the aforementioned algorithms
in the three-part and four-part splitting methods for other
choices of parameters and initial conditions. Taking $E=0.995$,
$L=4$, $\beta=1\times 10^{-3}$ and $r=15.5$, we obtain a
figure-eight orbit on the Poincar\'{e} section in Figure 6(a).
This figure-eight orbit has a hyperbolic fixed point, which
corresponds to a stable direction and another unstable direction.
It is a separation layer between the regular and chaotic regions.
The accuracies of the fourth-order integrators in Figure 6(b) are
similar to those in Figure 5(a). There are small differences
between Figures 6(b) and  5(a). The accuracy of each integrator in
Figure 6(b) is about one order higher than that in Figure 5(a).
The accuracies of the sixth-order integrators (such as
$PRK_{10}6B$) in Figure 6(c) have no explicit advantages over
those of the fourth-order integrators (e.g., $PRK_{6}4B$) in
Figure 6(b). When a chaotic orbit with parameters $E=0.992$,
$L=4$, $\beta=1.7\times 10^{-3}$ and initial separation $r=25$ in
Figure 7(a) is selected as a test orbit, the optimal fourth-order
method $PRK_{6}4B$ or $RKN_{6}4B$ in Figure 7(b) has several
orders of magnitude better in accuracies than the fourth-order
method S4A. The sixth-order methods in Figure 7(c) are not
explicitly superior to the fourth-order PRK and RKN methods in
Figure 7(b).

The main result can be concluded from Figures 5-7 and Tables 2 and
3. The optimal fourth-order methods $PRK_{6}4B$ and $RKN_64B$ are
the best ones of the aforementioned algorithms, and show the best
numerical performance in computational accuracy and efficiency.

\section{Conclusions}

Explicit symplectic integrators are not available for curved
spacetimes such as the Schwarzschild or Kerr type spacetimes if
the Hamiltonians corresponding to these spacetimes are split into
two parts like Hamiltonian problems in the solar system. This is
because the two parts lack the separation of variables, are
nonintegrable, or have analytical solutions which are not explicit
functions of time but are implicit functions of time. A series of
recent works (Wang et al. 2021a, 2021b, 2021c; Wu et al. 2021)
have successfully worked out this obstacle. The basic idea is
splitting the considered Hamiltonians or time-transformed
Hamiltonianss into more parts whose analytical solutions are
explicit functions of time.

A notable point is that the Hamiltonian splitting method is not
unique but has various choices. Taking a Hamiltonian describing
the motion of charged particles around the Schwarzschild  black
hole immersed in an external magnetic field as an example, we can
easily separate the Hamiltonian into three, four and five
explicitly integrable parts, which are expected to have analytical
solutions as explicit functions of time. Errors of an integrator
of order 2 or 4 closely depend on the Hamiltonian splitting
method. Given an appropriate time step, this integrator shows the
best accuracy in the three-part splitting method, but the poorest
accuracy in the five-part splitting method. This result is
independent of the type of orbits which are regular or chaotic.

It is also found that the optimized  fourth-order PRK and RKN
explicit symplectic integration schemes in the three-part
splitting are several orders of magnitude better in accuracies
than the fourth-order Yoshida methods. The former algorithms need
small additional computational cost compared with the latter ones.
The optimized  sixth-order PRK and RKN explicit symplectic
integrators have no dramatic advantages over the optimized
fourth-order ones in accuracies during long-term integrations due
to the rapid accumulation of roundoff errors.

Although the choice of the best explicit symplectic integrators is
based on the Schwarzschild spacetime backgrounds, it is applicable
to the Kerr type spacetimes or other curved spacetimes. That is,
time-transformed Hamiltonians associated with the Kerr type
spacetimes, or (time-transformed) Hamiltonians corresponding to
the other curved spacetimes, should decrease the number of
explicitly integrable splitting parts. Such a splitting method is
helpful to decrease the number of computations and then reduces
the roundoff errors. In this case, the optimized  fourth-order PRK
and RKN explicit symplectic integrators will exhibit the best
performance.

\appendix
\section*{APPENDIX}
\section{Extended phase space methods}

The  four-dimensional phase space Hamiltonian in Equation (6) is
labelled as $H=H(r,\theta ,p_{r},p_{\theta})$. Following the idea
of Pihajoki (2015) and extending the phase space to an
eight-dimensional phase space $(r,\theta
,\tilde{r},\tilde{\theta}, p_{r}, p_{\theta },
\tilde{p}_{r},\tilde{p}_{\theta})$, we have a new Hamiltonian
$\Gamma$ as follows:
\begin{eqnarray}
  \Gamma &=& \Gamma_{1}+\Gamma_{2},  \\
\Gamma_{1} &=& H(r, \theta ,
   \tilde{p}_{r},\tilde{p}_{\theta}), \nonumber \\
\Gamma_{2} &=& H(\tilde{r},\tilde{\theta},
   p_{r}, p_{\theta}). \nonumber
\end{eqnarray}
Clearly, $\Gamma_1$ and $\Gamma_2$ are independently analytically
solvable. Advancing time $h$, their flows are described by
$\Xi^{\Gamma_1}_{h}$ and $\Xi^{\Gamma_2}_{h}$. Equation (13)
becomes
\begin{eqnarray}
   S2\Gamma(h)=\Xi^{\Gamma_1}_{h/2}\times \Xi^{\Gamma_2}_{h}\times
   \Xi^{\Gamma_1}_{h/2}.
\end{eqnarray}
Replacing S2A with  S2$\Gamma$ in Equation (14), we obtain a
fourth-order explicit symplectic integrator  S4$\Gamma$ in the
extended phase space. Luo et al. (2017) introduced a midpoint
permutation matrix
\begin{equation}
  \Psi=
   \left[
      \begin{array}{cccccccccc}
        \frac{1}{2} &0 &\frac{1}{2}&0 &0&0&0&0 \\
        0 &\frac{1}{2}&0& \frac{1}{2} &0&0&0&0\\
        \frac{1}{2} &0 &\frac{1}{2}&0 &0&0&0&0\\
        0 &\frac{1}{2}&0& \frac{1}{2} &0&0&0&0\\
        0 &0&0&0 &\frac{1}{2}&0& \frac{1}{2} &0 \\
        0 &0&0&0 &0&\frac{1}{2}& 0 &\frac{1}{2} \\
        0 &0&0&0 &\frac{1}{2}&0& \frac{1}{2} &0 \\
        0 &0&0&0 &0&\frac{1}{2}& 0 &\frac{1}{2} \\
      \end{array}
   \right].
\end{equation}
In fact, this matrix means the following transformations
\begin{eqnarray}
  r &=& \tilde{r}\leftarrow \frac{r+\tilde{r}}{2}, ~~~~~~~~~~ \theta=\tilde{\theta}\leftarrow
  \frac{\theta+\tilde{\theta}}{2}, \\
   p_r &=& \tilde{p}_{r}\leftarrow \frac{p_r+\tilde{p}_{r}}{2}, ~~~~ p_\theta=\tilde{p}_\theta\leftarrow
  \frac{p_\theta+\tilde{p}_\theta}{2}.
\end{eqnarray}
The method S4$\Gamma$ combining the matrix $\Psi$ corresponds to a
fourth-order explicit scheme
\begin{eqnarray}
   LS4=\Psi \times S4\Gamma.
\end{eqnarray}
Due to the inclusion of the permutation $\Psi$, LS4 is a
symplectic-like method for the extended phase space Hamiltonian
$\Gamma$. Similarly, Equations (16) and (17) become
\begin{eqnarray}
   \chi_{\Gamma h} =
   \Xi_{h}^{\Gamma_{2}} \times \Xi_{h}^{\Gamma_{1}}, ~~~~
   \chi_{\Gamma h}^{*} = \Xi_{h}^{\Gamma_{1}}
     \times \Xi_{h}^{\Gamma_{2}}.
\end{eqnarray}
Using $\chi_{\Gamma}$ and $\chi_{\Gamma}^{*}$ instead of
$\chi_{A}$ and $\chi_{A}^{*}$ in Equation (23), we have
$PRK_64\Gamma$. Thus, an extended phase space PRK explicit
symplectic-like method is
\begin{eqnarray}
   LPRK_64=\Psi \times PRK_64\Gamma.
\end{eqnarray}

Adding a third part
\begin{eqnarray}
  \Gamma_3 = \frac{\omega}{2}[(r-\tilde{r})^{2}+(
\theta-\tilde{\theta})^{2}+(p_{r}-\tilde{p}_{r})^{2}+ (
p_{\theta}-\tilde{p}_{\theta})^{2}]
\end{eqnarray}
to Equation (A1), Tao (2016a) obtained another new Hamiltonian
\begin{eqnarray}
K=\Gamma +\Gamma_{3}.
\end{eqnarray}
Here, $\omega$ is a parameter controlling the binding of the two
copies. Noting that $H_1\rightarrow \Gamma_{1}$,
$\mathcal{H}_2\rightarrow \Gamma_{2}$ and $H_3\rightarrow
\Gamma_{3}$ in Equation (28), we have TS4 corresponding to S4B in
Equation (36) and TPRK$_6$4 corresponding to PRK$_6$4B. TS4 and
TPRK$_6$4 are two fourth-order extended phase space explicit
symplectic methods for the Hamiltonian $K$. Numerical accuracies
depend on the control parameter $\omega$. For $\omega=3.553$ in
Figure 8(a), the method TS4 has the best accuracy. For
$\omega=7.943$ in Figure 8(b), the method TPRK$_6$4 performs the
best accuracy. The two values of $\omega$ are considered in other
computations.

\section*{Acknowledgments}

The authors are very grateful to the anonymous referee for
valuable comments and suggestions  that have greatly improved this
article. This research has been supported by the National Natural
Science Foundation of China (Grant Nos. 11973020 and 11533004) and
the National Natural Science Foundation of Guangxi (No.
2019JJD110006).

 \begin{table*}
      \begin{center}
      \small \caption{Coefficients of optimal explicit symplectic PRK and RKN methods}
      \begin{tabular}{ccc}\hline
      $PRK_{6}4$: order 4, ~~ $s=6$\\
      $\alpha_{1}=\alpha_{12}=0.079203696431196$ & $\alpha_{2}=\alpha_{11}=0.130311410182166$ & $\alpha_{3}=\alpha_{10}=0.222861495867608$ \\
      $\alpha_{4}=\alpha_{9}=-0.366713268047426$ & $\alpha_{5}=\alpha_{8}=0.324648188689706$ & $\alpha_{6}=\alpha_{7}=0.109688477876750$ \\
      \\
     $RKN_{6}4$: order 4, ~~ $s=6$\\
      $\alpha_{1}=\alpha_{12}=0.082984402775764$ & $\alpha_{2}=\alpha_{11}=0.162314549088478$ & $\alpha_{3}=\alpha_{10}=0.233995243906975$ \\
      $\alpha_{4}=\alpha_{9}=0.370877400040627$ & $\alpha_{5}=\alpha_{8}=-0.409933704882860$ & $\alpha_{6}=\alpha_{7}=0.059762109071016$ \\
      \hline
     $PRK_{10}6$:  order 6, ~~ $s=10$\\
      $\alpha_{1}=\alpha_{20}=0.050262764400392$ & $\alpha_{2}=\alpha_{19}=0.098553687334061$ & $\alpha_{3}=\alpha_{18}=0.314960598945618$ \\
      $\alpha_{4}=\alpha_{17}=-0.447346463799477$ & $\alpha_{5}=\alpha_{16}=0.492426354438066$ & $\alpha_{6}=\alpha_{15}=-0.425118748098612$ \\
      $\alpha_{7}=\alpha_{14}=0.237063888460398$ & $\alpha_{8}=\alpha_{13}=0.195602502673864$ & $\alpha_{9}=\alpha_{12}=0.346358153969049$ \\
      $\alpha_{10}=\alpha_{11}=-0.362762738019228$ \\
      \\
     $RKN_{11}6$: order 6, ~~ $s=11$\\
      $\alpha_{1}=\alpha_{22}=0.041464999318123$ & $\alpha_{2}=\alpha_{21}=0.081764779984951$ & $\alpha_{3}=\alpha_{20}=0.116363890469074$ \\
      $\alpha_{4}=\alpha_{19}=0.174189917743206$ & $\alpha_{5}=\alpha_{18}=-0.214196108281612$ & $\alpha_{6}=\alpha_{17}=0.087146900594235$ \\
      $\alpha_{7}=\alpha_{16}=-0.011892914772034$ & $\alpha_{8}=\alpha_{15}=-0.234438851475716$ & $\alpha_{9}=\alpha_{14}=0.222927464172244$ \\
      $\alpha_{10}=\alpha_{13}=0.134281413629651$ & $\alpha_{11}=\alpha_{12}=0.102388508617878$ \\
      \\
     $RKN_{14}6$: order 6, ~~ $s=14$\\
      $\alpha_{1}=\alpha_{28}=0.037859320640564$ & $\alpha_{2}=\alpha_{27}=0.053859829902649$ & $\alpha_{3}=\alpha_{26}=0.048775799572468$ \\
      $\alpha_{4}=\alpha_{25}=0.135207377374172$ & $\alpha_{5}=\alpha_{24}=-0.161075266078115$ & $\alpha_{6}=\alpha_{23}=0.104540901258588$ \\
      $\alpha_{7}=\alpha_{22}=0.209700508043170$ & $\alpha_{8}=\alpha_{21}=-0.204785819165409$ & $\alpha_{9}=\alpha_{20}=0.074641357176006$ \\
      $\alpha_{10}=\alpha_{19}=0.069119771011173$ & $\alpha_{11}=\alpha_{18}=0.037297929637134$ & $\alpha_{12}=\alpha_{17}=0.291269754059613$ \\
      $\alpha_{13}=\alpha_{16}=-0.3000639975070951$ & $\alpha_{14}=\alpha_{15}=0.103652534075081$ \\
      \hline
      \end{tabular} 
   \end{center}
   \end{table*}

\begin{table*}
   \begin{center}
   \small \caption{CPU times (unit: minute$'$ second$''$ ) for all  integrators acting on the three
   splitting  methods considered in Figures 3, 4 and 5. The test orbit is the regular orbit in Figure 1,
   and the integration time of each algorithm reaches $\tau =10^{7}$.}
      \begin{tabular}{ccccccc}
           \hline
      Algorithm & S2A & S2B & S2C & S4A & S4B & S4C \\
      \hline
      CPU Time & 0$'$32$''$ & 1$'$13$''$ & 1$'$36$''$ & 1$'$03$''$ & 1$'$22$''$ & 1$'$47$''$ \\
      \hline
      Algorithm & S6A & S6B & S6C & $PRK_{6}4A$ & $PRK_{6}4B$ & $PRK_{6}4C$ \\
      \hline
      CPU Time & 2$'$17$''$ & 2$'$18$''$ & 2$'$31$''$ & 1$'$48$''$ & 1$'$53$''$ & 2$'$11$''$ \\
      \hline
      Algorithm & $RKN_{6}4A$ & $RKN_{6}4B$ & $RKN_{6}4C$ & $PRK_{10}6A$ & $PRK_{10}6B$ & $PRK_{10}6C$ \\
      \hline
      CPU Time & 2$'$0$''$ & 2$'$12$''$ & 2$'$31$''$ & 2$'$29$''$ & 2$'$37$''$ & 2$'$57$''$ \\
      \hline
     Algorithm & $RKN_{11}6A$ & $RKN_{11}6B$ & $RKN_{11}6C$ & $RKN_{14}6A$ & $RKN_{14}6B$ & $RKN_{14}6C$ \\
      \hline
      CPU Time & 2$'$35$''$ & 2$'$43$''$ & 3$'$13$''$ & 2$'$47$''$ & 2$'$5$''$ & 3$'$23$''$  \\
      \hline
      Algorithm & $TS4$ & $LS4$ & $TPRK_{6}4$ & $LPRK_{6}4$ \\
      \hline
      CPU Time & 1$'$12$''$ & 0$'$41$''$ & 2$'$27$''$ & 1$'$55$''$ \\
      \hline
         \end{tabular} 
   \end{center}
   \end{table*}

    \begin{table*}
      \begin{center}
      \small \caption{The minimum and maximum energy errors for all integrators in Figure 5.}
      \begin{tabular}{cccccccccccc}\hline
Algorithm & Minimum & Maximum &  Algorithm & Minimum & Maximum \\
\hline
S4A & $10^{-10.86}$ & $10^{-8.09}$ & S4B & $10^{-11.56}$ & $10^{-8.79}$  \\
\hline S6A & $10^{-13.02}$ & $10^{-10.61}$ &  S6B & $10^{-13.67}$
& $10^{-11.52}$ \\
\hline
$PRK_{6}4A$ & $10^{-13.64}$ & $10^{-10.79}$ & $PRK_{6}4B$ & $10^{-14.78}$ & $10^{-12.12}$ \\
\hline
 $RKN_{6}4A$ & $10^{-12.50}$ & $10^{-10.55}$ & $RKN_{6}4B$ & $10^{-13.83}$ &
 $10^{-11.88}$ \\ \hline
 $PRK_{10}6A$ & $10^{-15.15}$ & $10^{-12.28}$ & $PRK_{10}6B$ & $10^{-15.55}$ & $10^{-12.06}$
 \\ \hline
 $RKN_{11}6A$ & $10^{-13.81}$ & $10^{-12.32}$ & $RKN_{11}6B$ & $10^{-14.81}$ &
 $10^{-12.08}$ \\ \hline
 $RKN_{14}6A$ & $10^{-13.01}$ & $10^{-11.22}$ & $RKN_{14}6B$ & $10^{-13.69}$ & $10^{-11.79}$ \\
      \hline
 $TS4$ & $10^{-10.98}$ & $10^{-8.04}$ & $LS4$ & $10^{-10.67}$ &
 $10^{-8.54}$ \\ \hline
 $TPRK_{6}4$ & $10^{-12.93}$ & $10^{-10.51}$ & $LPRK_{6}4$ & $10^{-13.36}$ & $10^{-10.96}$ \\
           \hline
      \end{tabular} 
   \end{center}
   \end{table*}

\begin{figure*}
      \centering{
          \includegraphics[width=22pc]{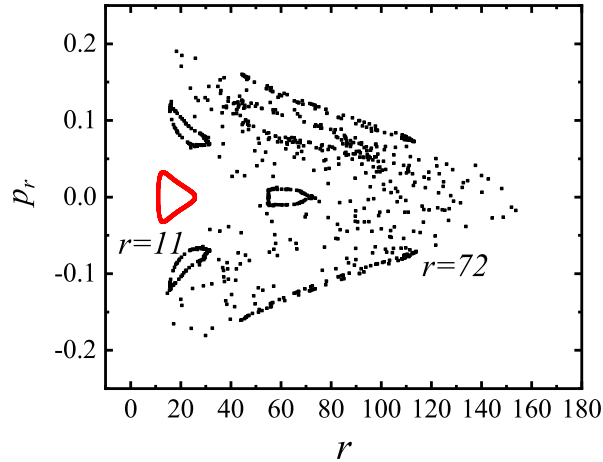}
          \caption{Poincar\'{e}
surface of section $\theta=\pi/2$ with $p_{\theta}<0$. The time
step uses $h=1$, and the parameters are $E=0.995$, $L=4.6$, and
$\beta=8.9\times10^{-4}$. The initial conditions are
$\theta=\pi/2$ and $p_{r}=0$; the initial value $p_{\theta}>0$ is
determined by Equation (7). The second-order method is applied to
the five-part splitting with $a=1$, i.e., S2C. A regular orbit
with initial separation $r=11$ colored Red and another chaotic
orbit with initial separation $r=72$ colored Black are plotted. }}
\end{figure*}

\begin{figure*}
\centering{
\includegraphics[width=18pc]{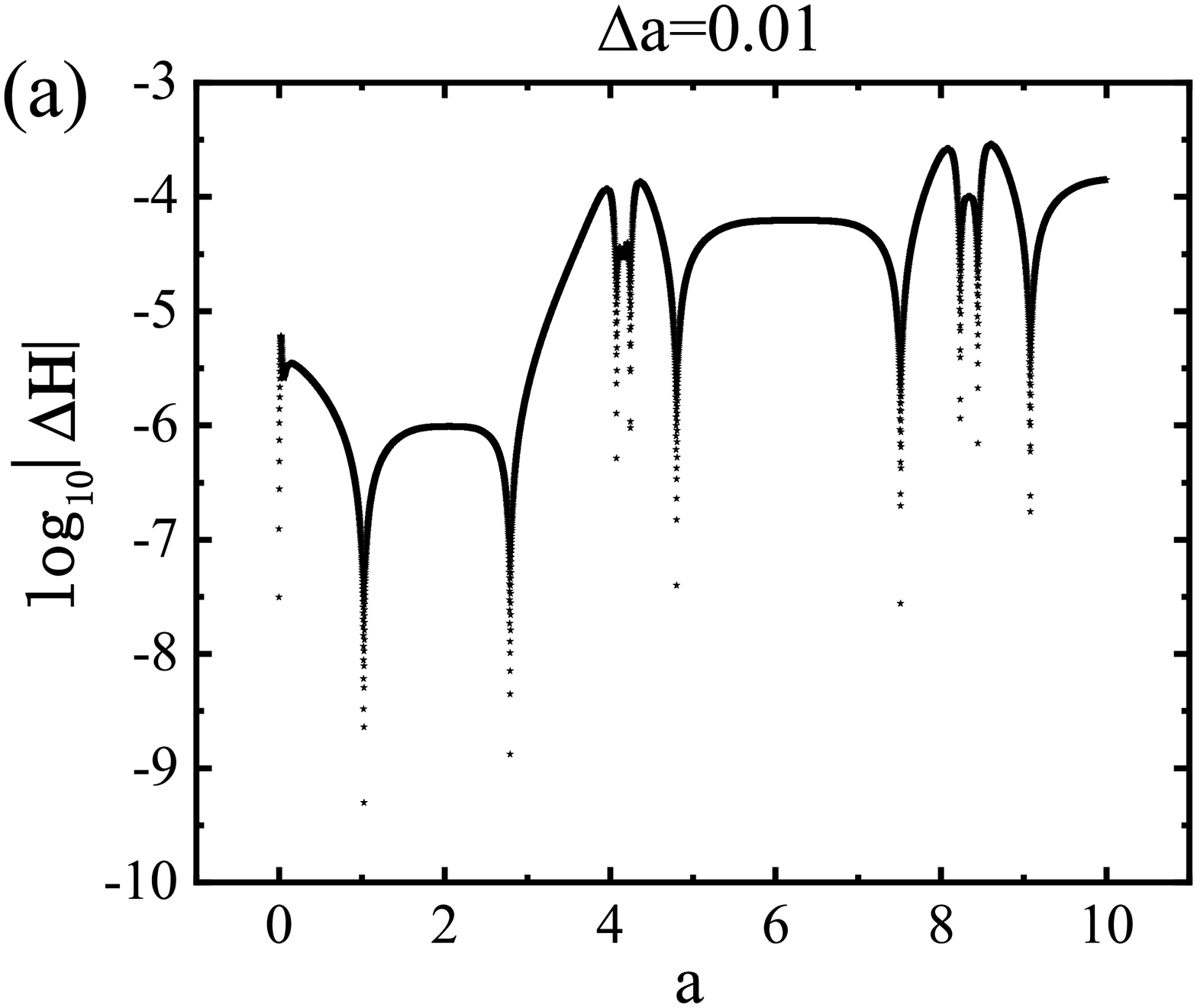}
\includegraphics[width=18pc]{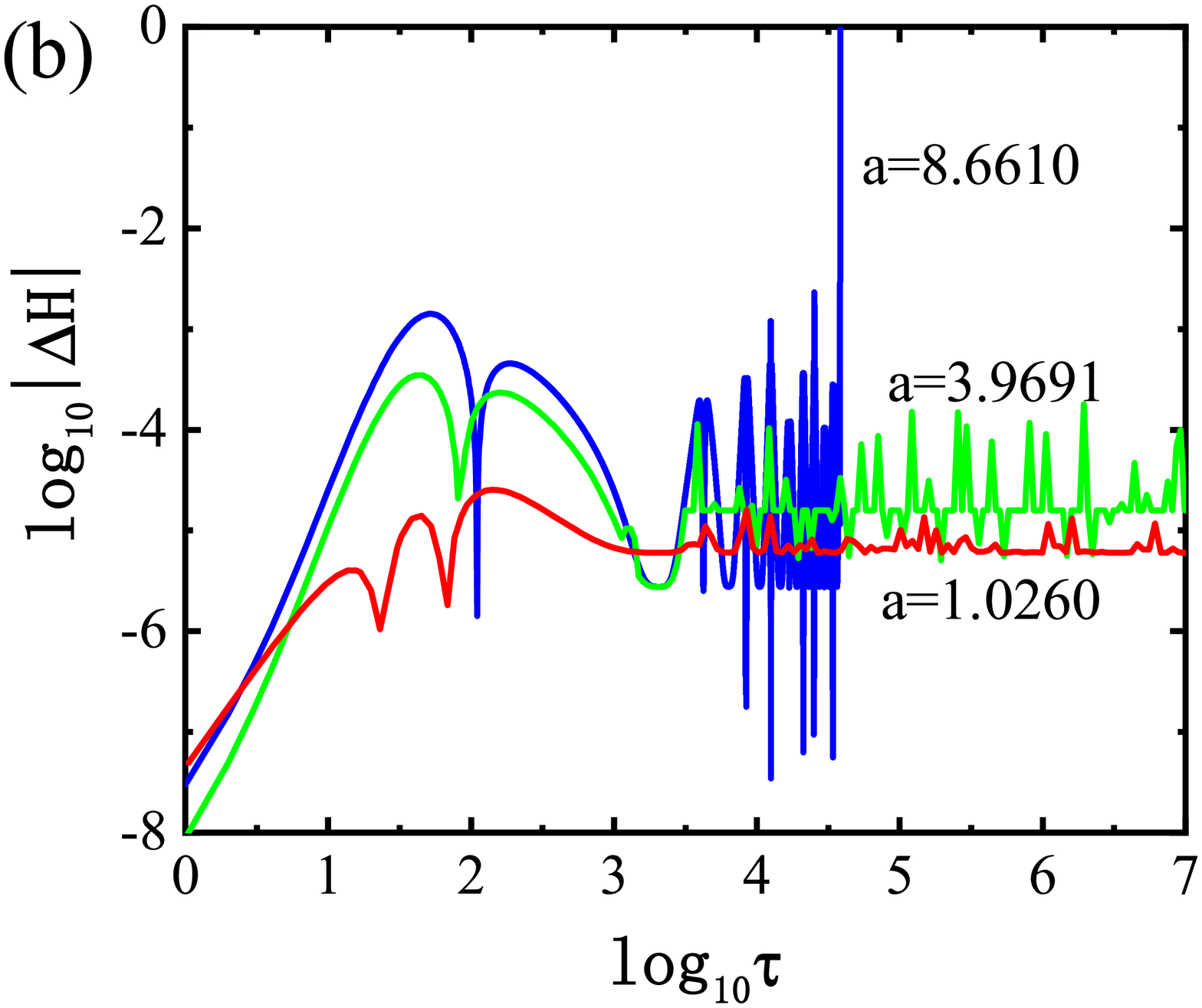}
\caption{(a) Dependence of Hamiltonian error $\Delta H=1+2H$ for
S2C solving the ordered orbit in Figure 1 on $a$. (b) Three values
of  $a$ in panel (a) correspond to Hamiltonian  errors. Clearly,
$a= 1.0260$ corresponds to the smallest error. It is considered in
later computations of the five-part splitting. } }
\end{figure*}

\begin{figure*}
\centering{
\includegraphics[width=12pc]{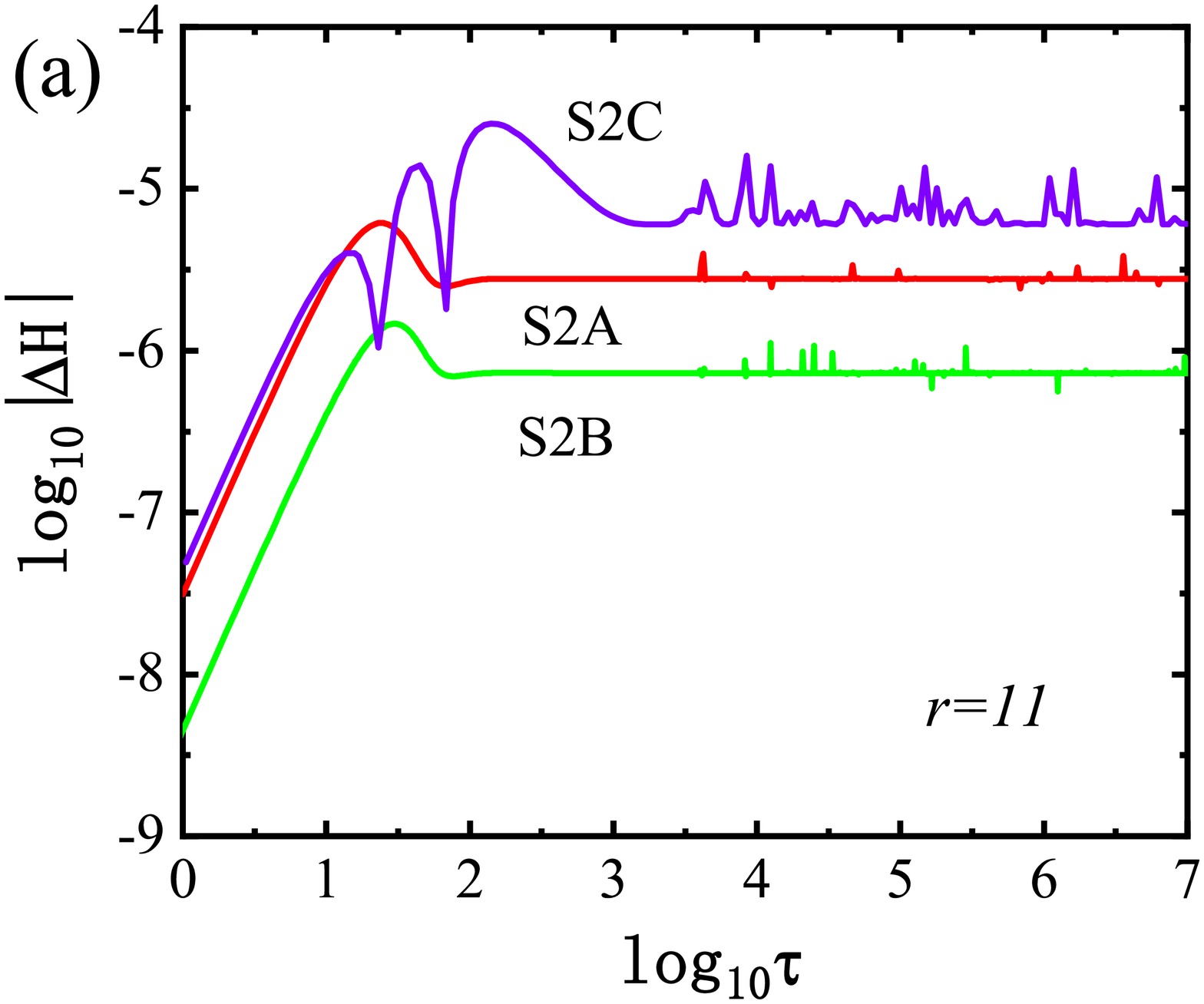}
\includegraphics[width=12pc]{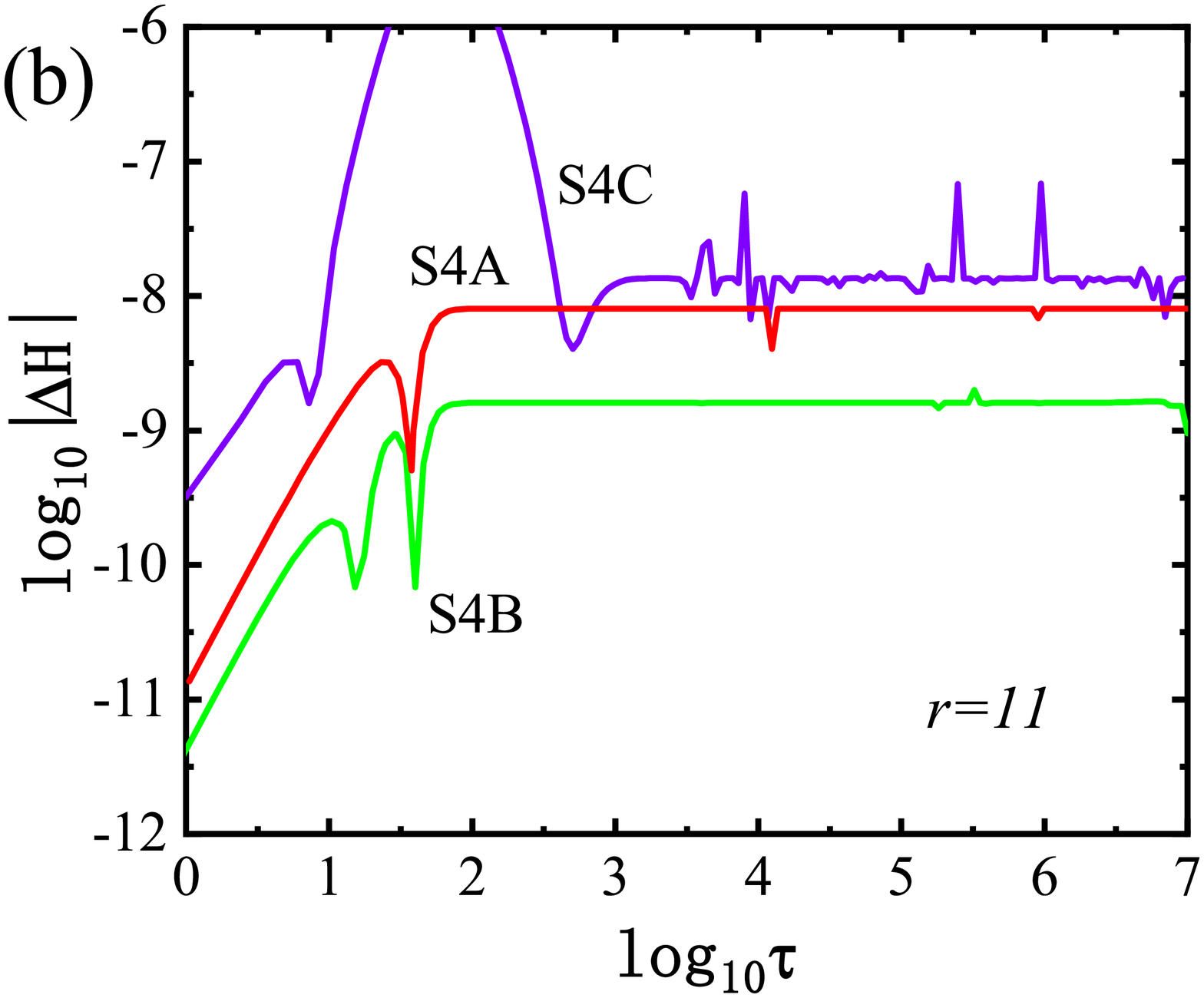}
\includegraphics[width=12pc]{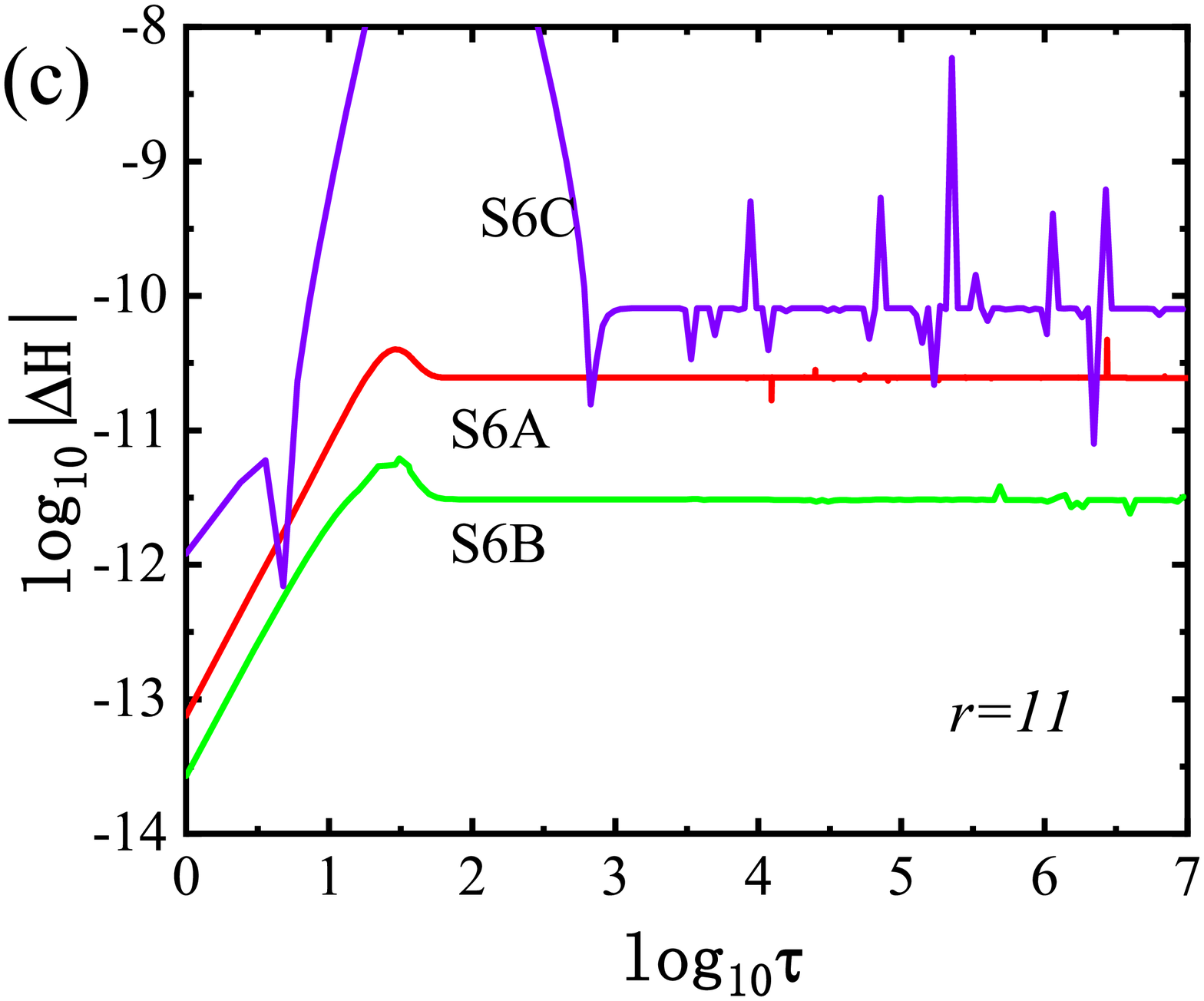}
\includegraphics[width=12pc]{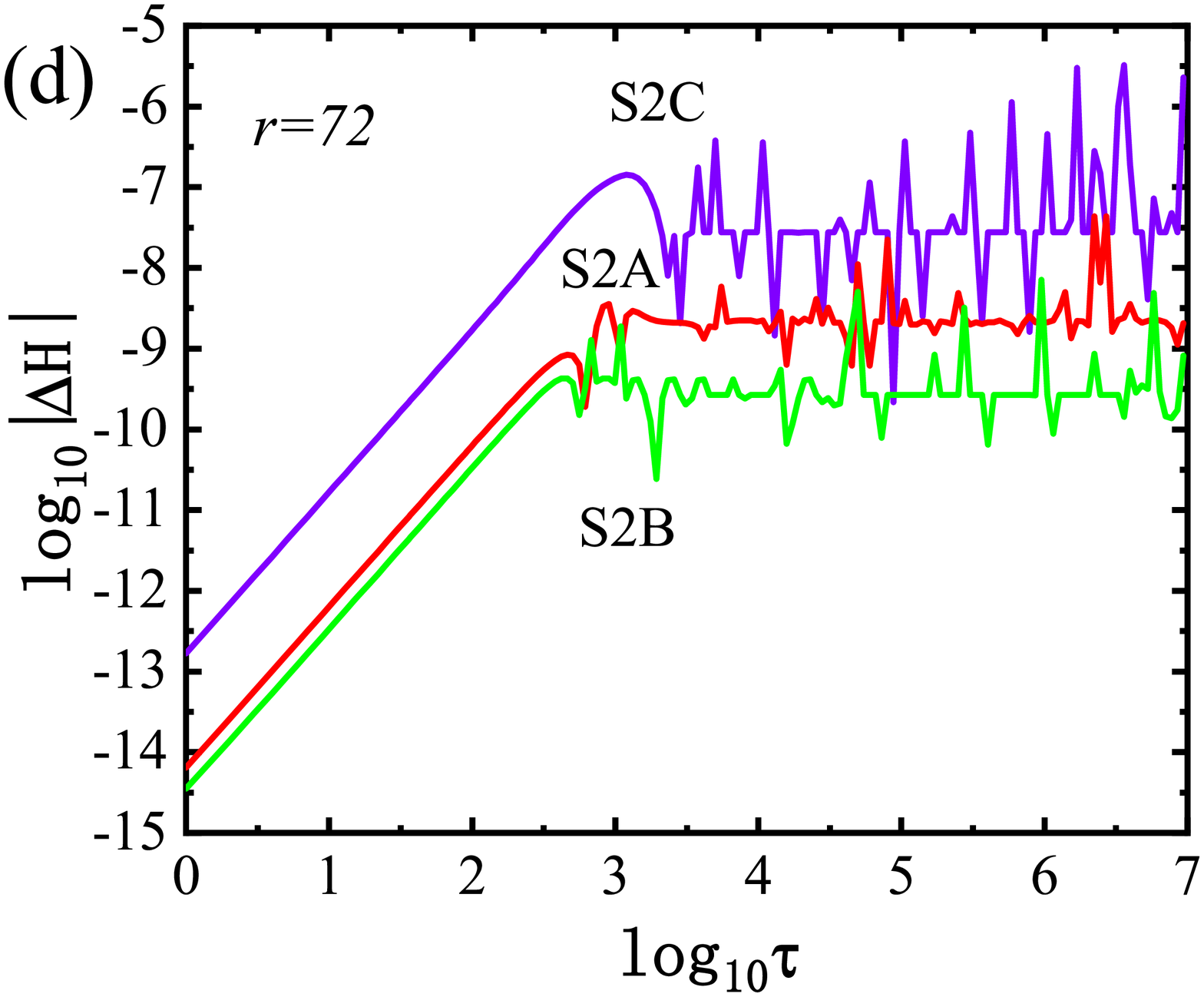}
\includegraphics[width=12pc]{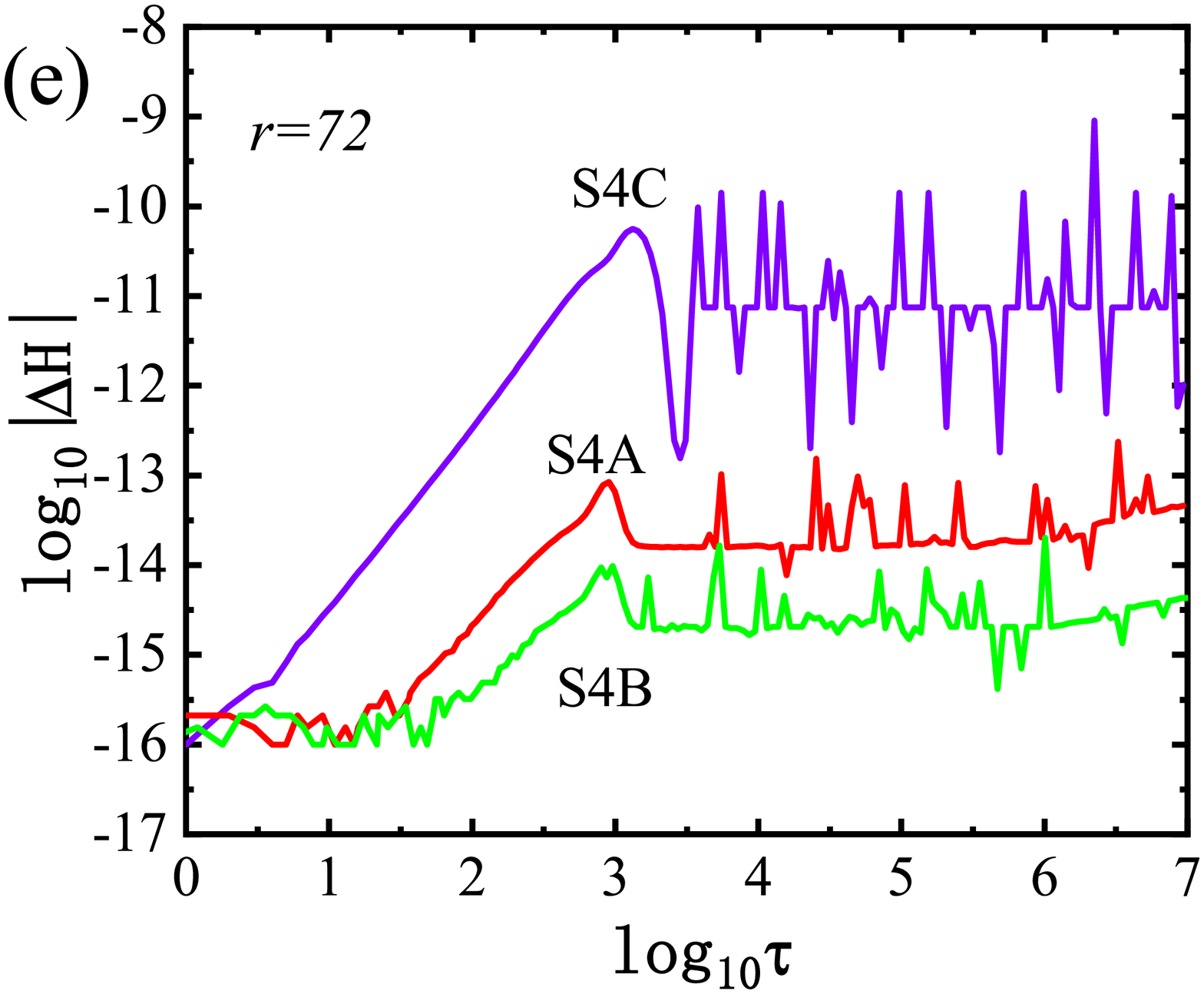}
\includegraphics[width=12pc]{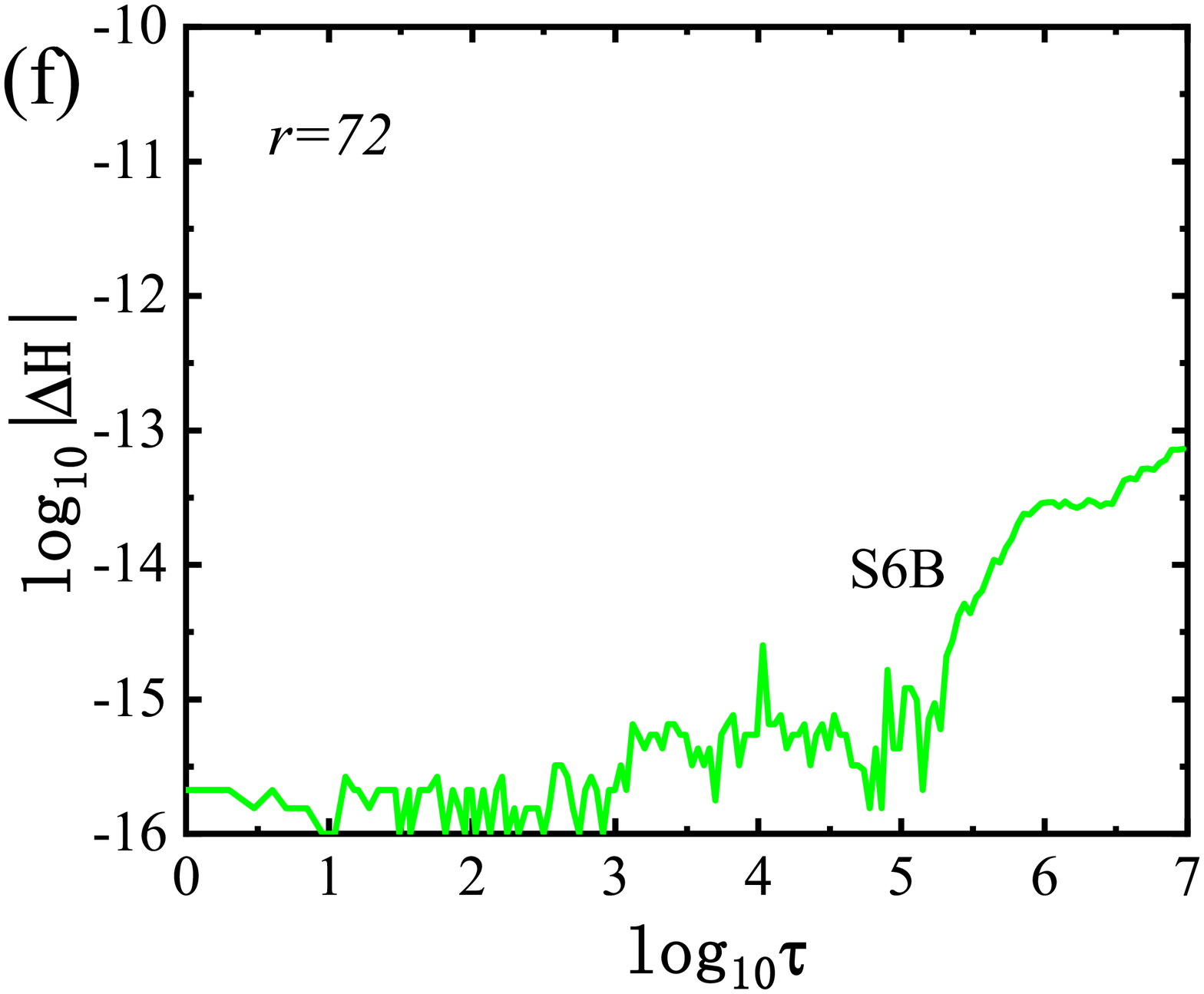}
\caption{(a-c):  Hamiltonian errors for the Yoshida type
integrators acting on the regular orbit in Figure 1. (d-f):
Similar to panels (a-c) but the regular orbit is replaced with the
chaotic orbit in Figure  1. S6A and S6C are almost consistent with
S6B. } }
\end{figure*}

\begin{figure*}
\centering{
\includegraphics[width=12pc]{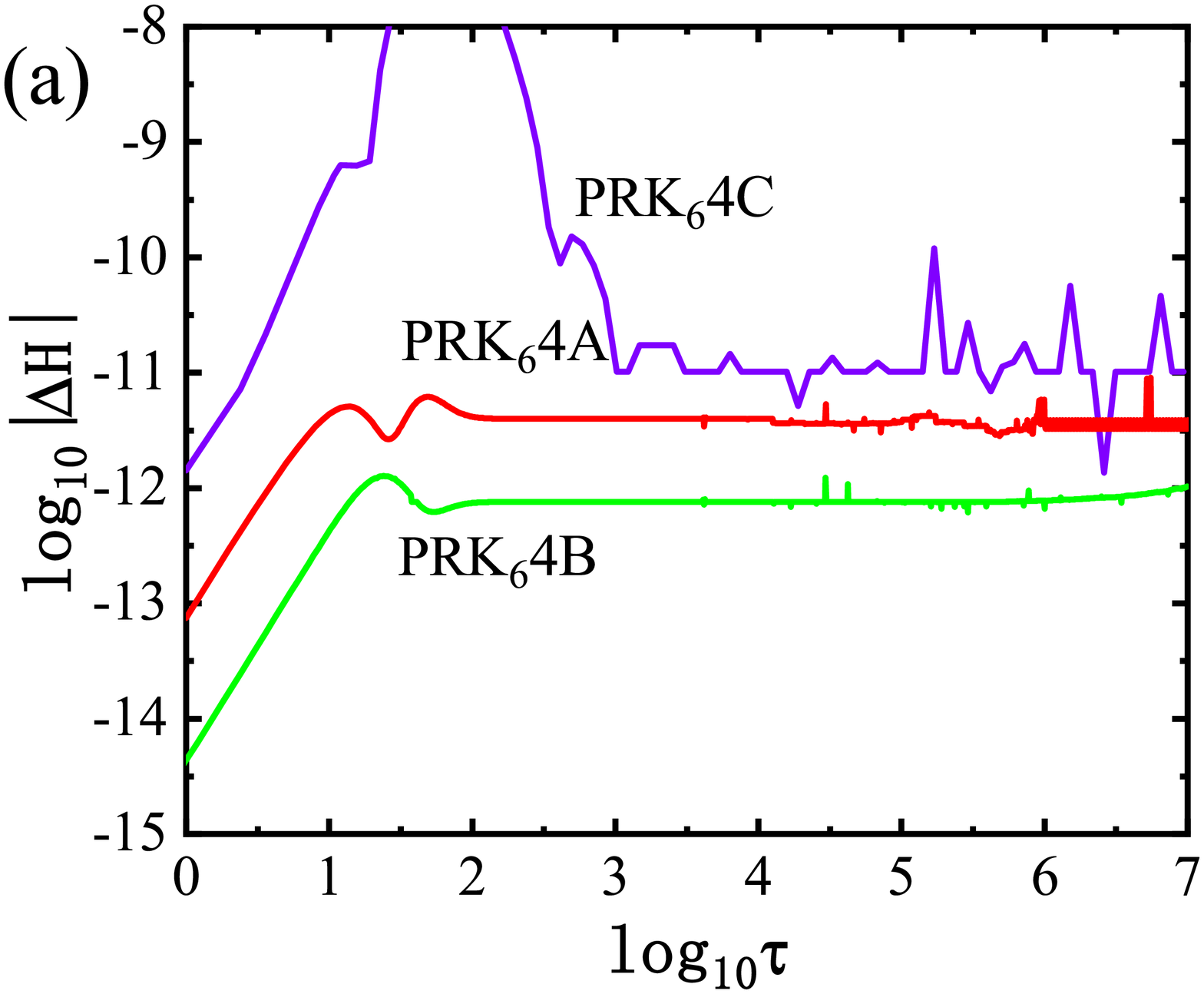}
\includegraphics[width=12pc]{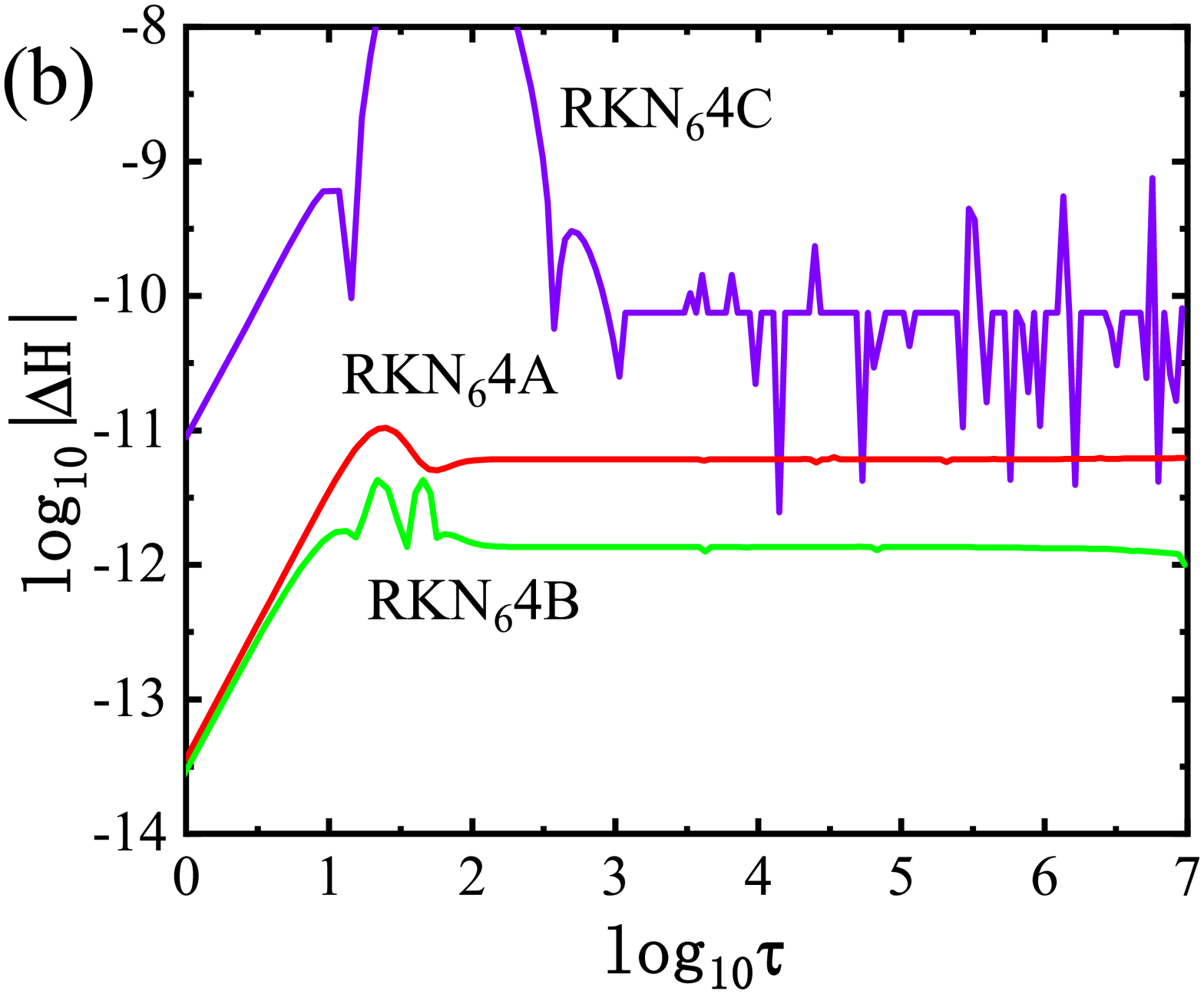}
\includegraphics[width=12pc]{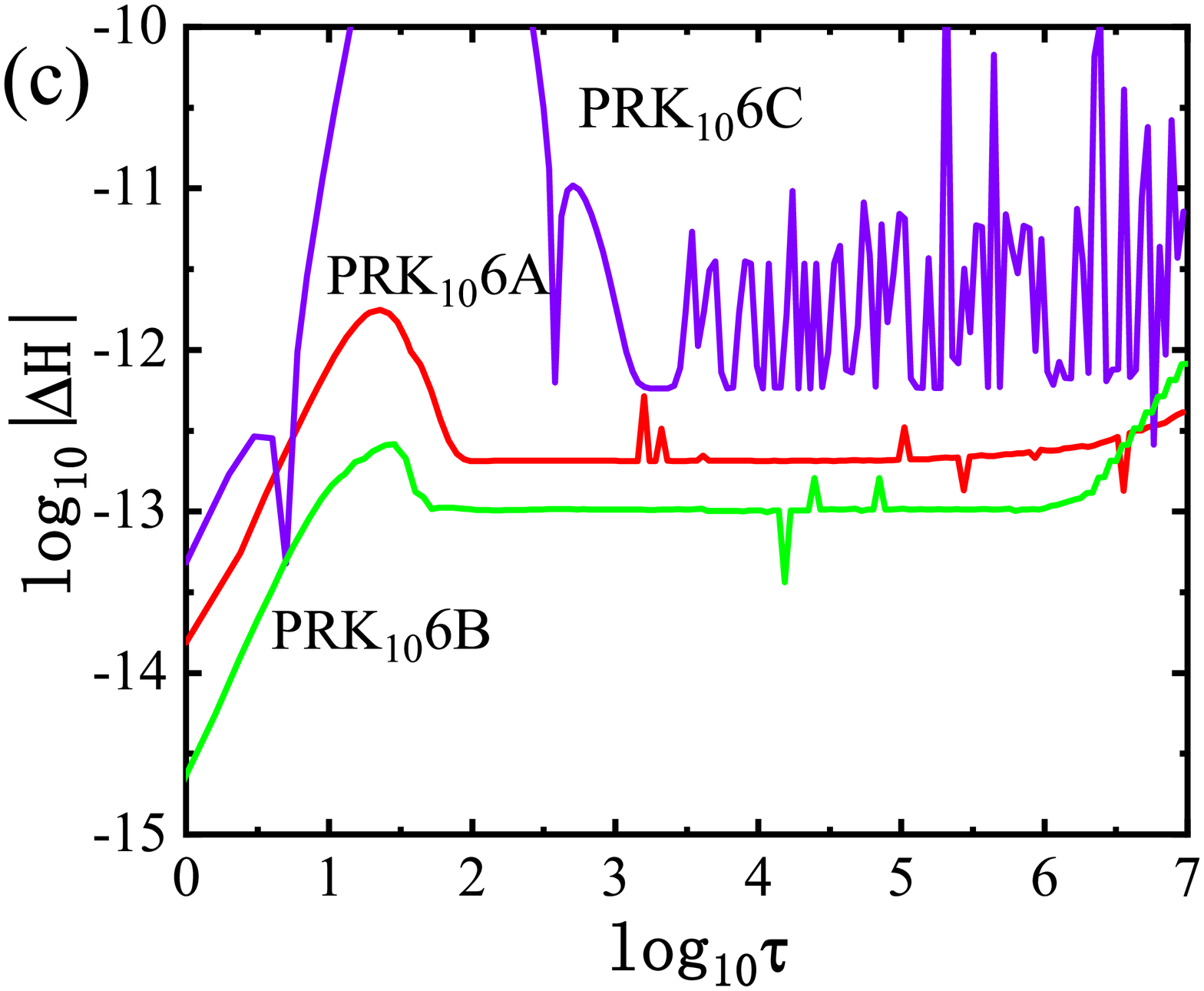}
\includegraphics[width=12pc]{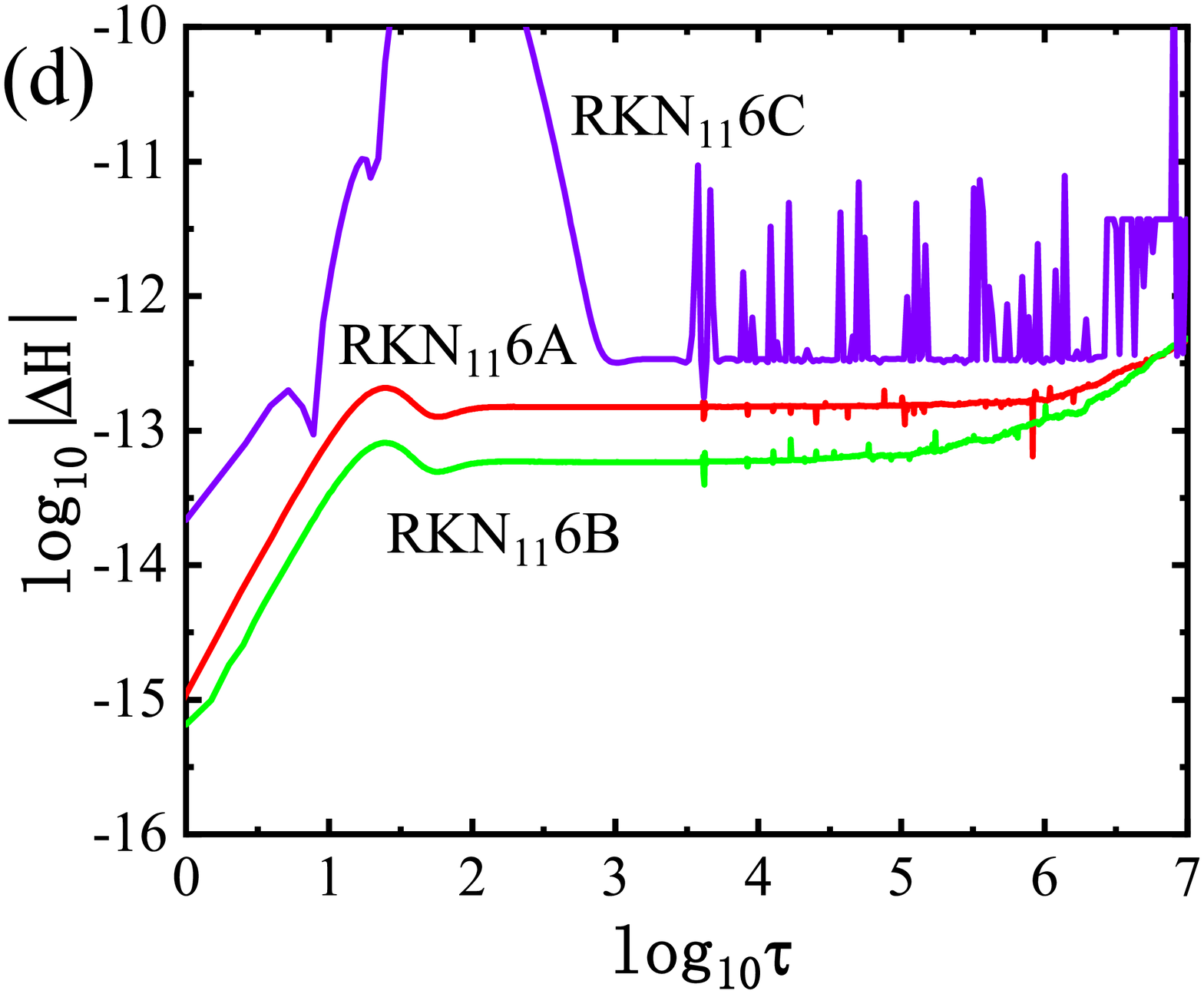}
\includegraphics[width=12pc]{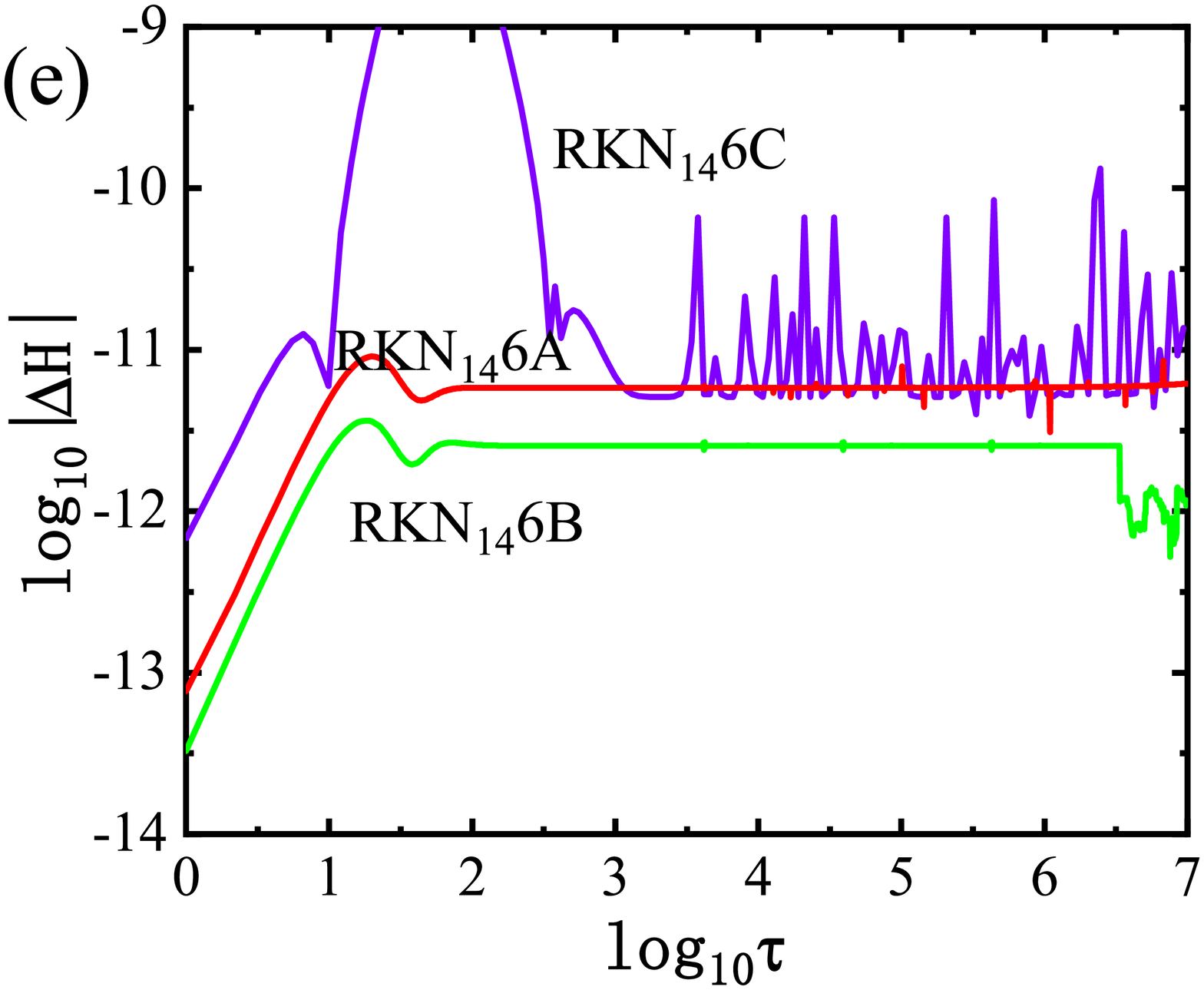}
\caption{Similar to Figures 3 (a-c) but the PRK and RKN
integrators are used in the three Hamiltonian splittings. (a)
Optimal fourth-order PRK methods. (b) Optimal fourth-order RKN
methods. (c) Optimal sixth-order PRK methods. (d) An optimal
sixth-order RKN method. (e) Another optimal sixth-order RKN
method.  } }
\end{figure*}

\begin{figure*}
\centering{
\includegraphics[width=12pc]{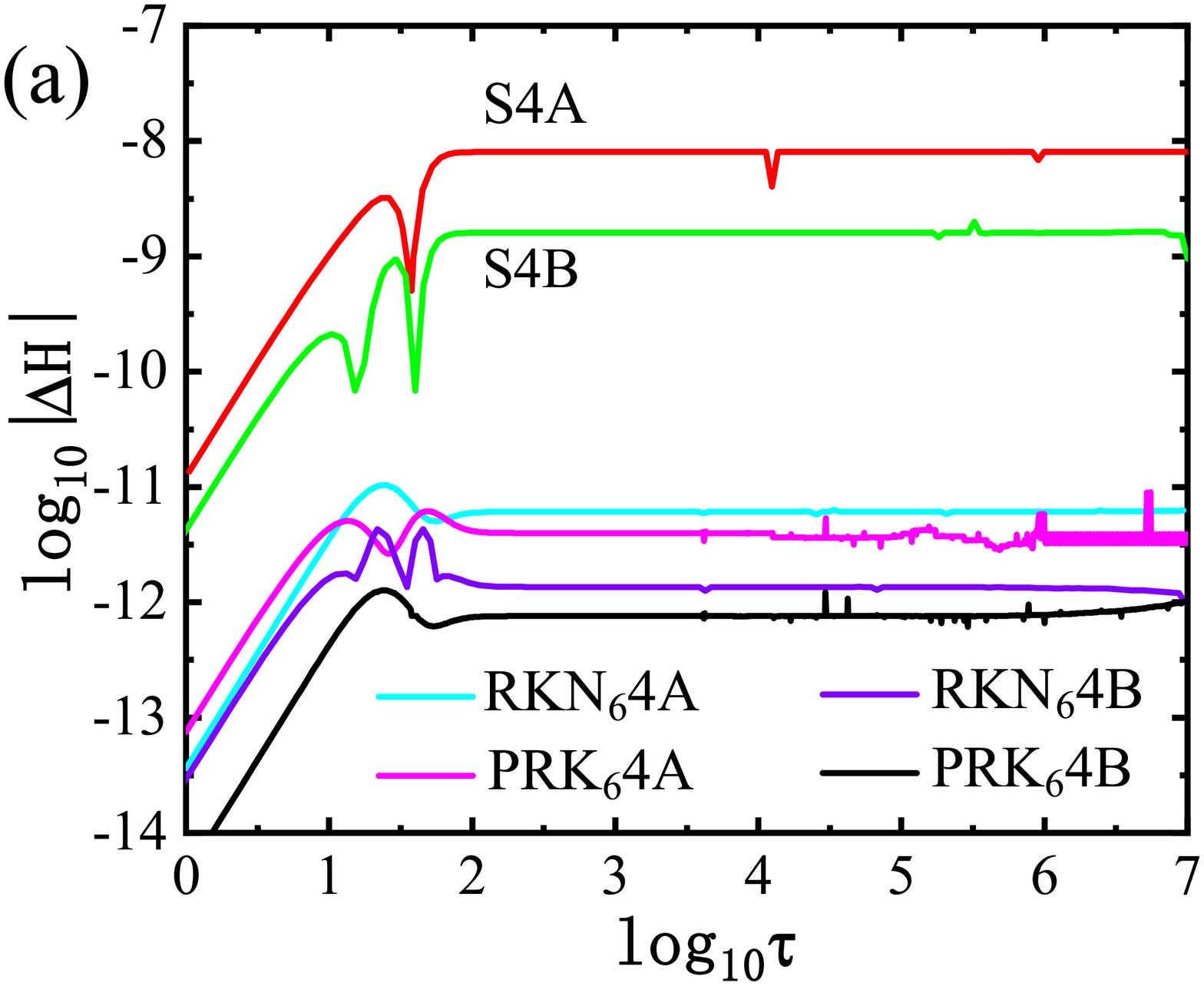}
\includegraphics[width=12pc]{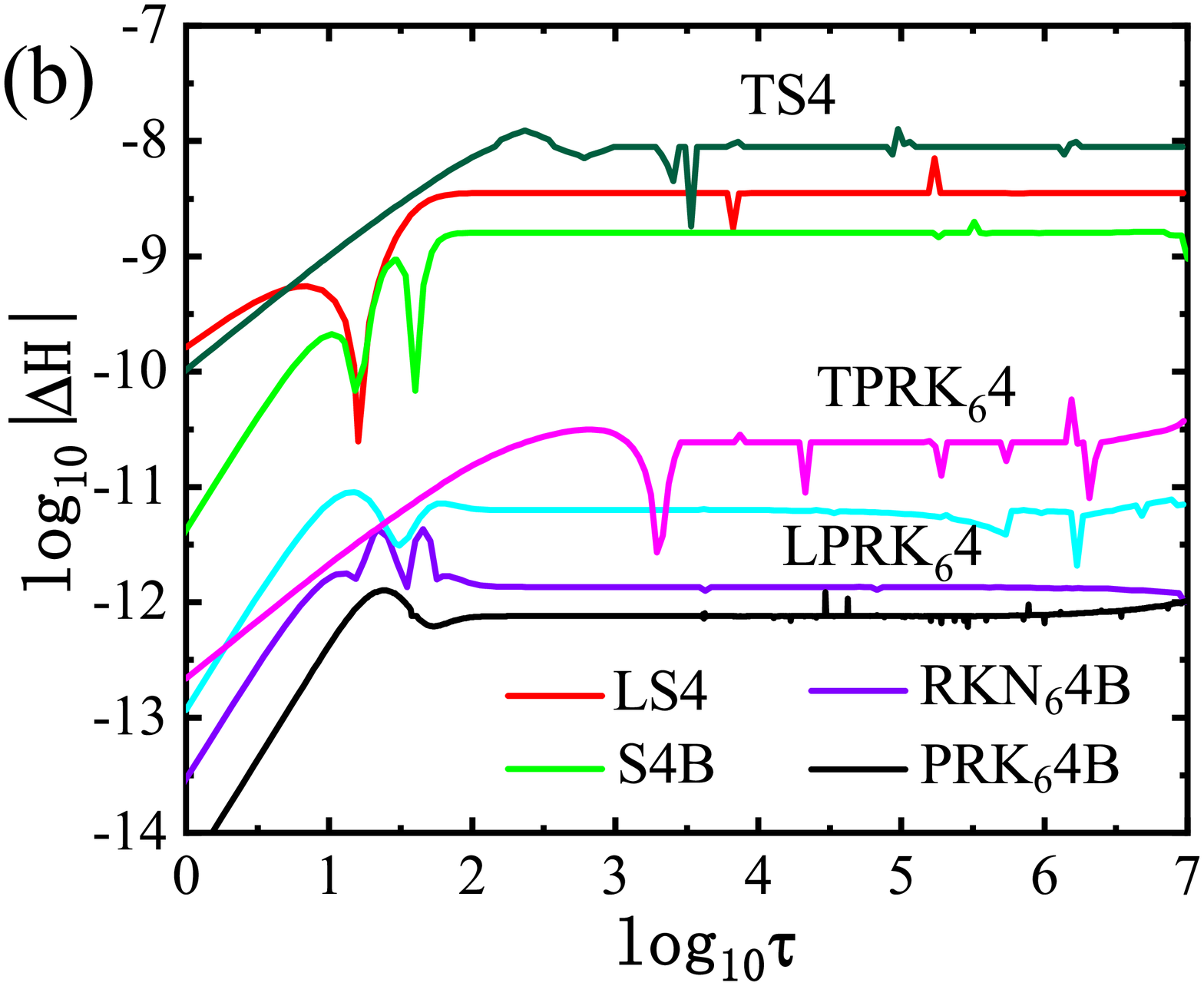}
\includegraphics[width=12pc]{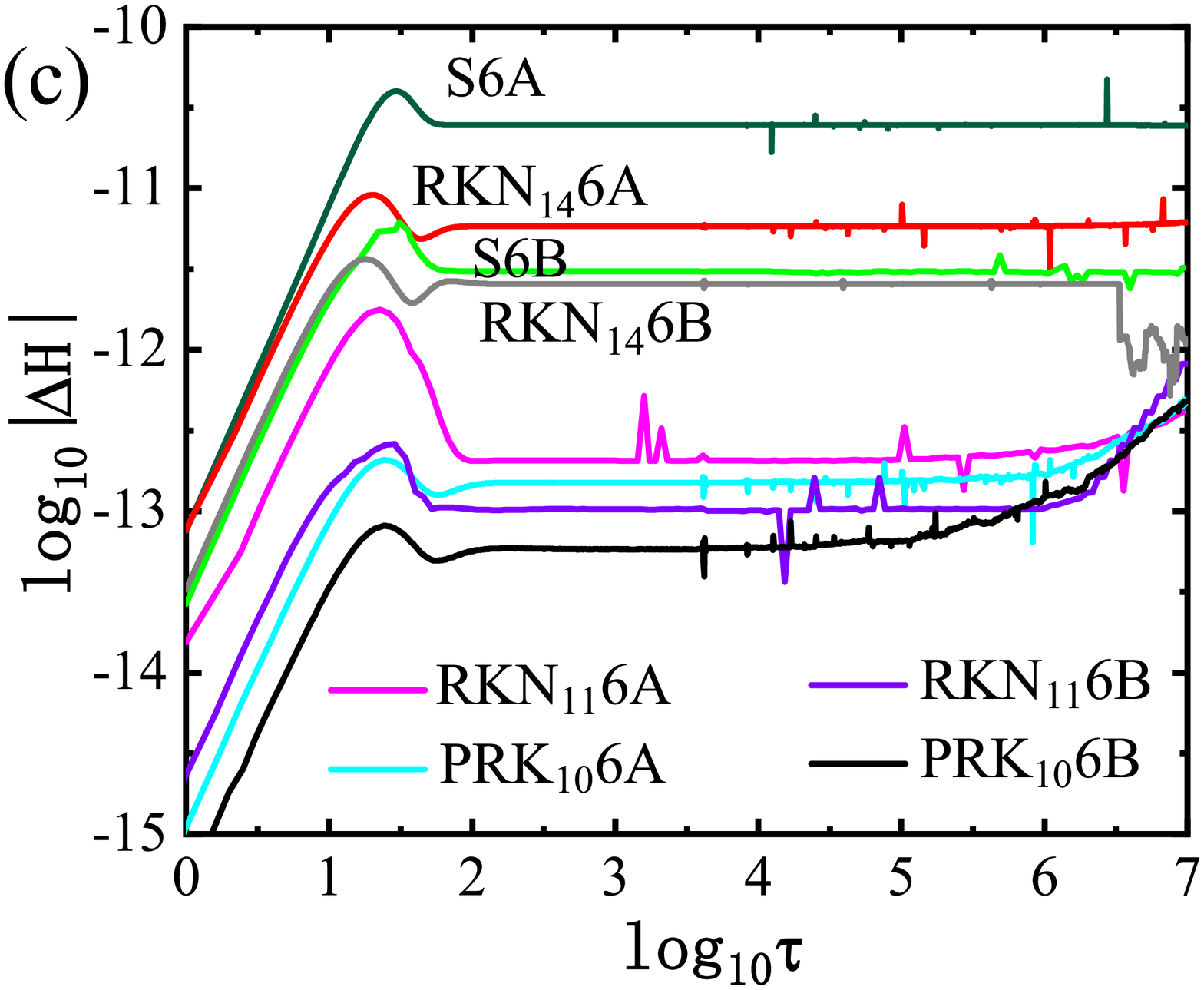}
\caption{Results concluded from those of the A and B type methods
in Figures 3 (a)-(e) and 4. (a) Errors of all fourth-order
integrators. (b) Errors of four methods TS4, LS4, TPRK$_{6}$4 and
LPRK$_{6}$4 in Appendix. } (c) Errors of all sixth-order
algorithms. }
\end{figure*}

\begin{figure*}
\centering{
\includegraphics[width=12pc]{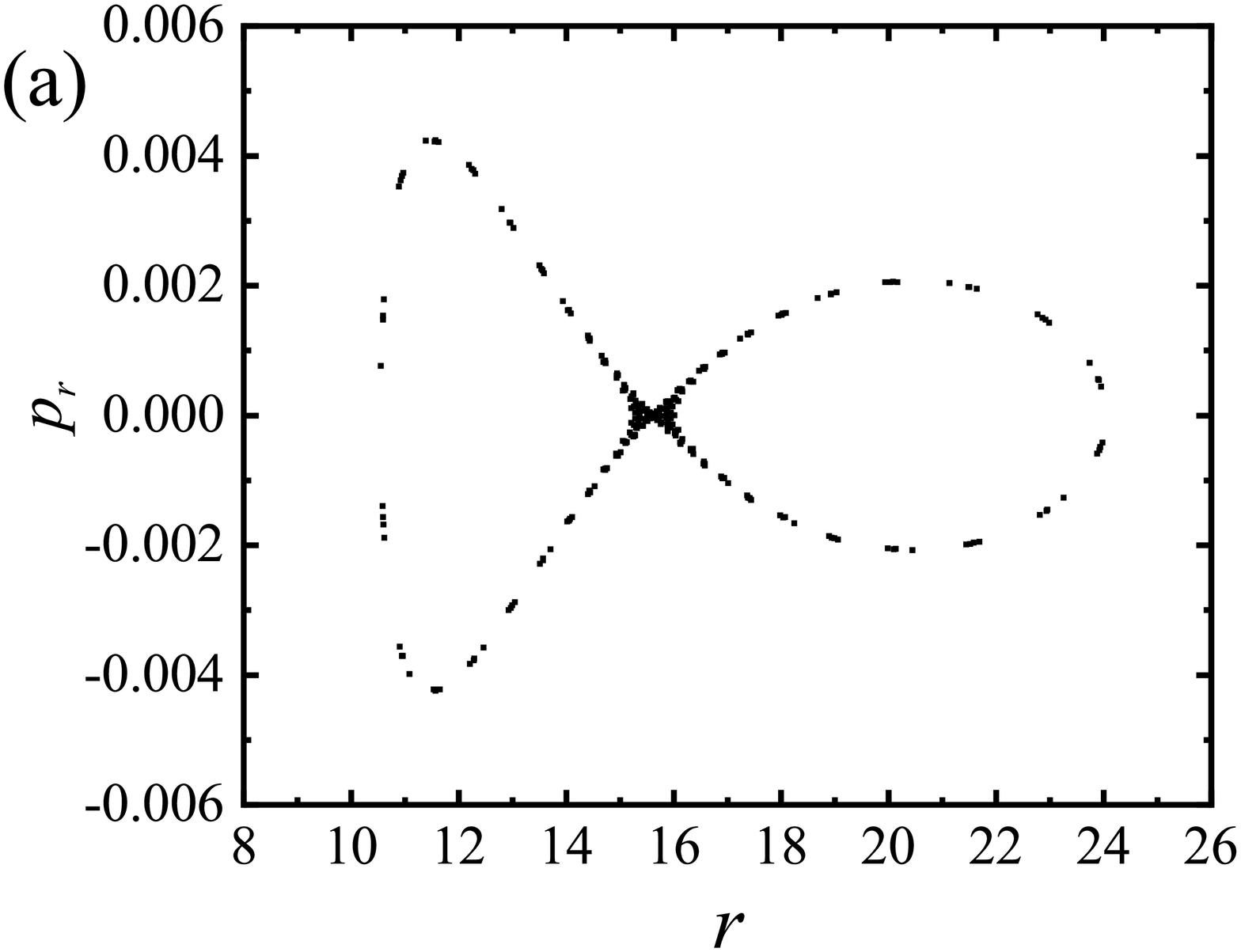}
\includegraphics[width=12pc]{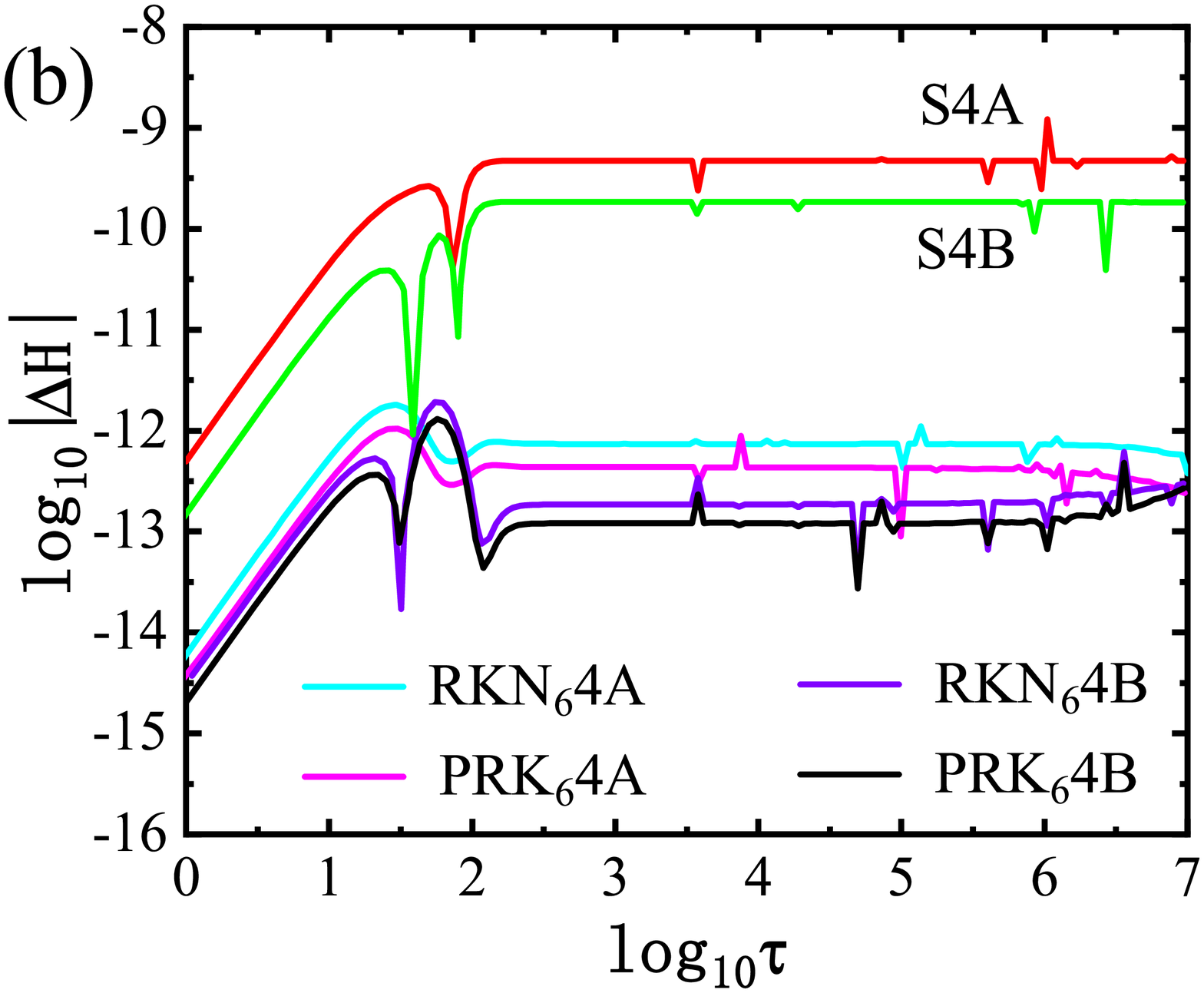}
\includegraphics[width=12pc]{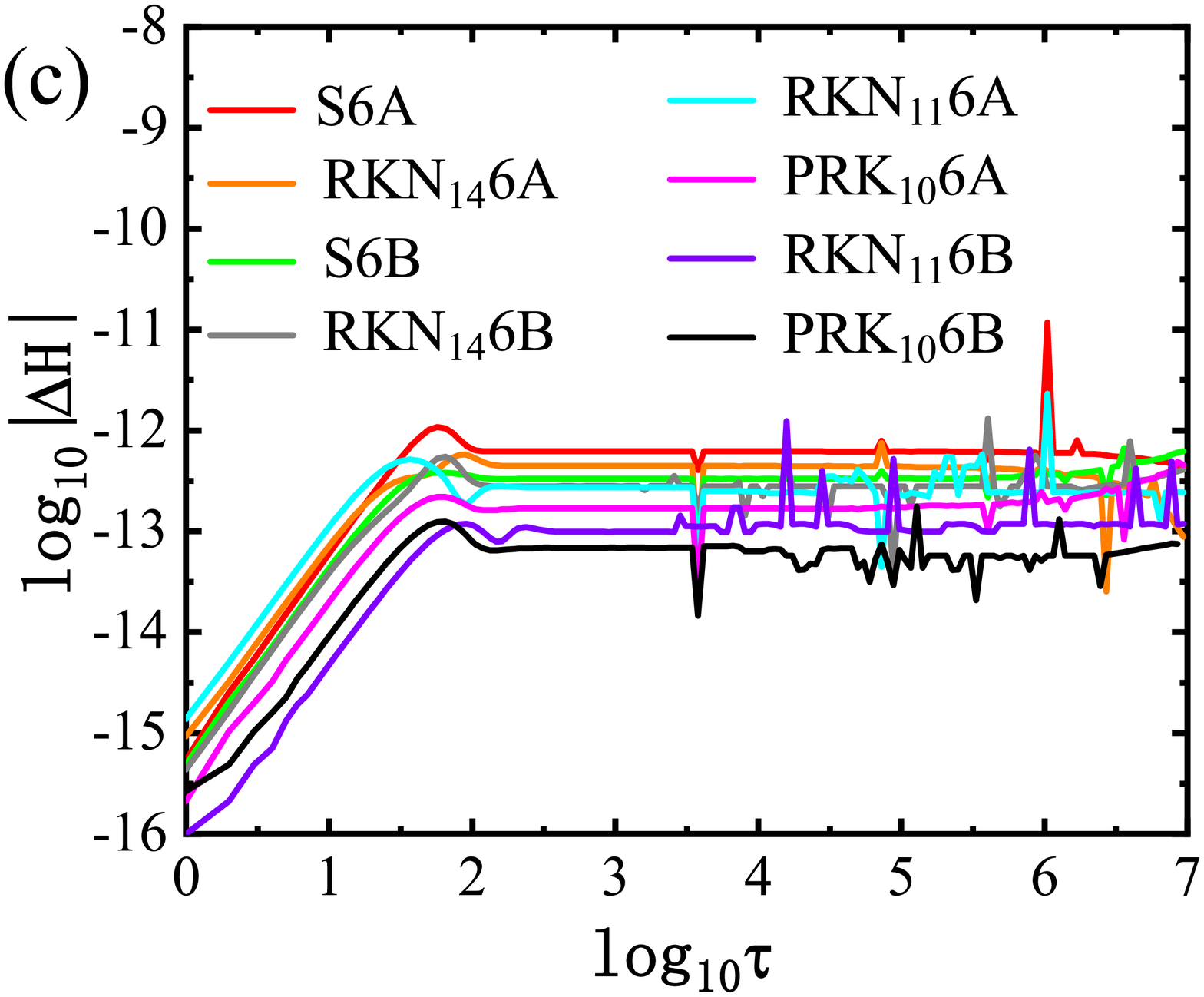}
\caption{(a) Figure-eight orbit with parameters $E=0.995$, $L=4$,
$\beta=1\times 10^{-3}$ and initial separation $r=15.5$ on the
Poincar\'{e} section. This orbit is described by $PRK_{6}4B$. (b)
Hamiltonian errors for all fourth-order A and B type integrators.
(c) Hamiltonian errors for all sixth-order A and B type
algorithms. }}
\end{figure*}

\begin{figure*}
\centering{
\includegraphics[width=12pc]{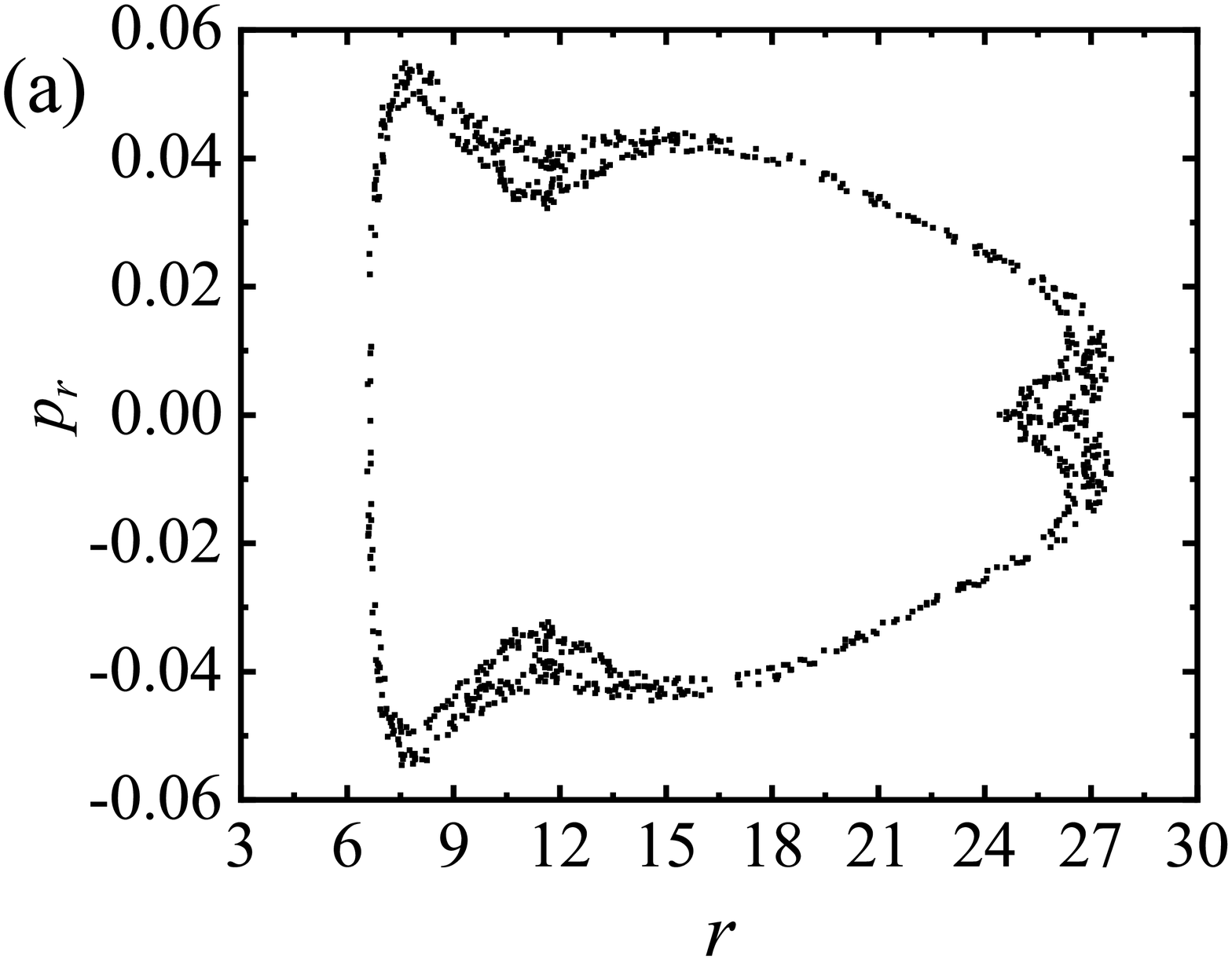}
\includegraphics[width=12pc]{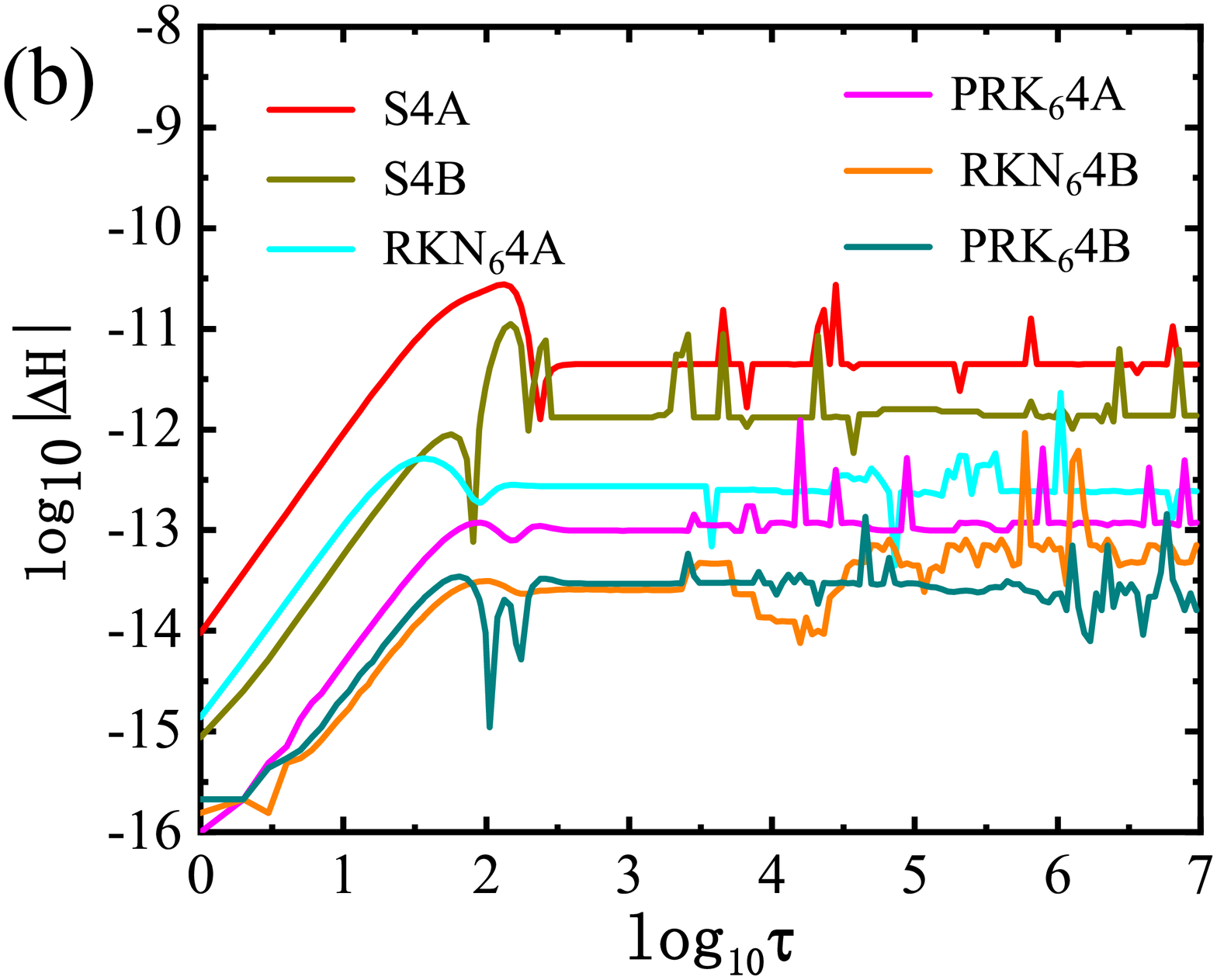}
\includegraphics[width=12pc]{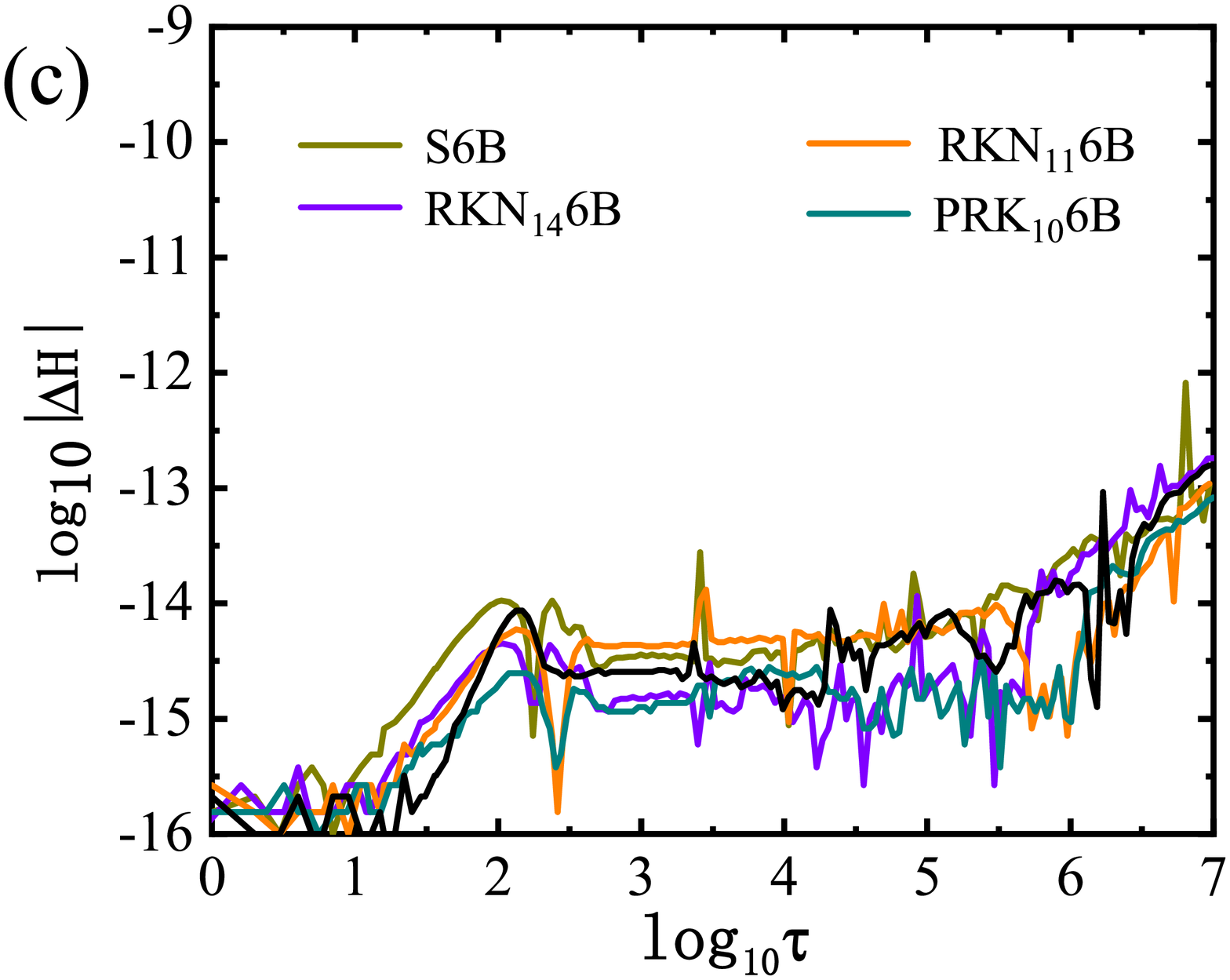}
\caption{(a) Chaotic orbit with parameters $E=0.992$, $L=4$,
$\beta=1.7\times 10^{-3}$ and initial separation $r=25$ on the
Poincar\'{e} section. This orbit is described by $PRK_{6}4B$. (b)
Hamiltonian errors for all fourth-order A and B type integrators.
(c) Hamiltonian errors for some sixth-order A and B type
algorithms. }}
\end{figure*}

\begin{figure*}
   \centering{
\includegraphics[width=18pc]{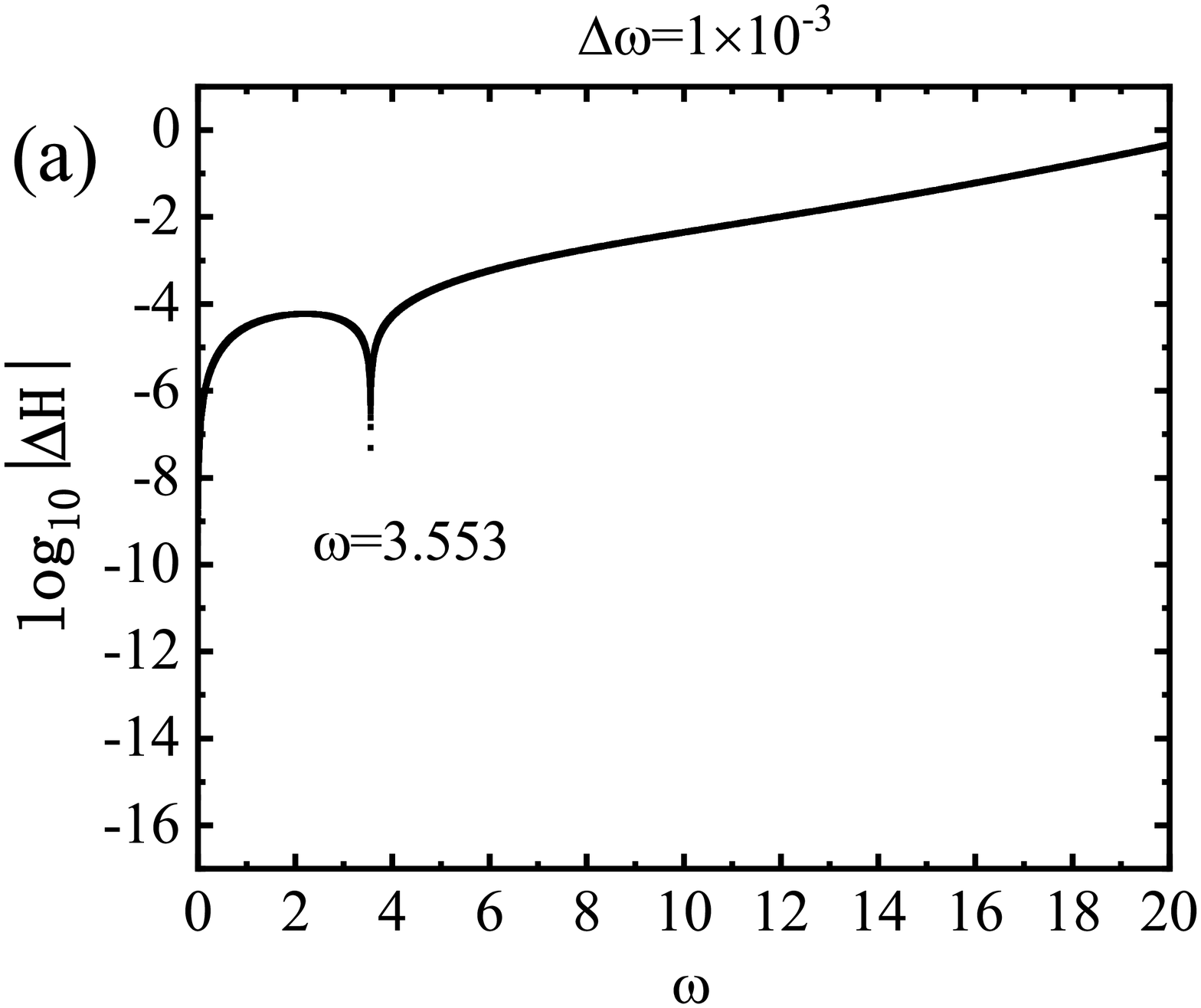}
\includegraphics[width=18pc]{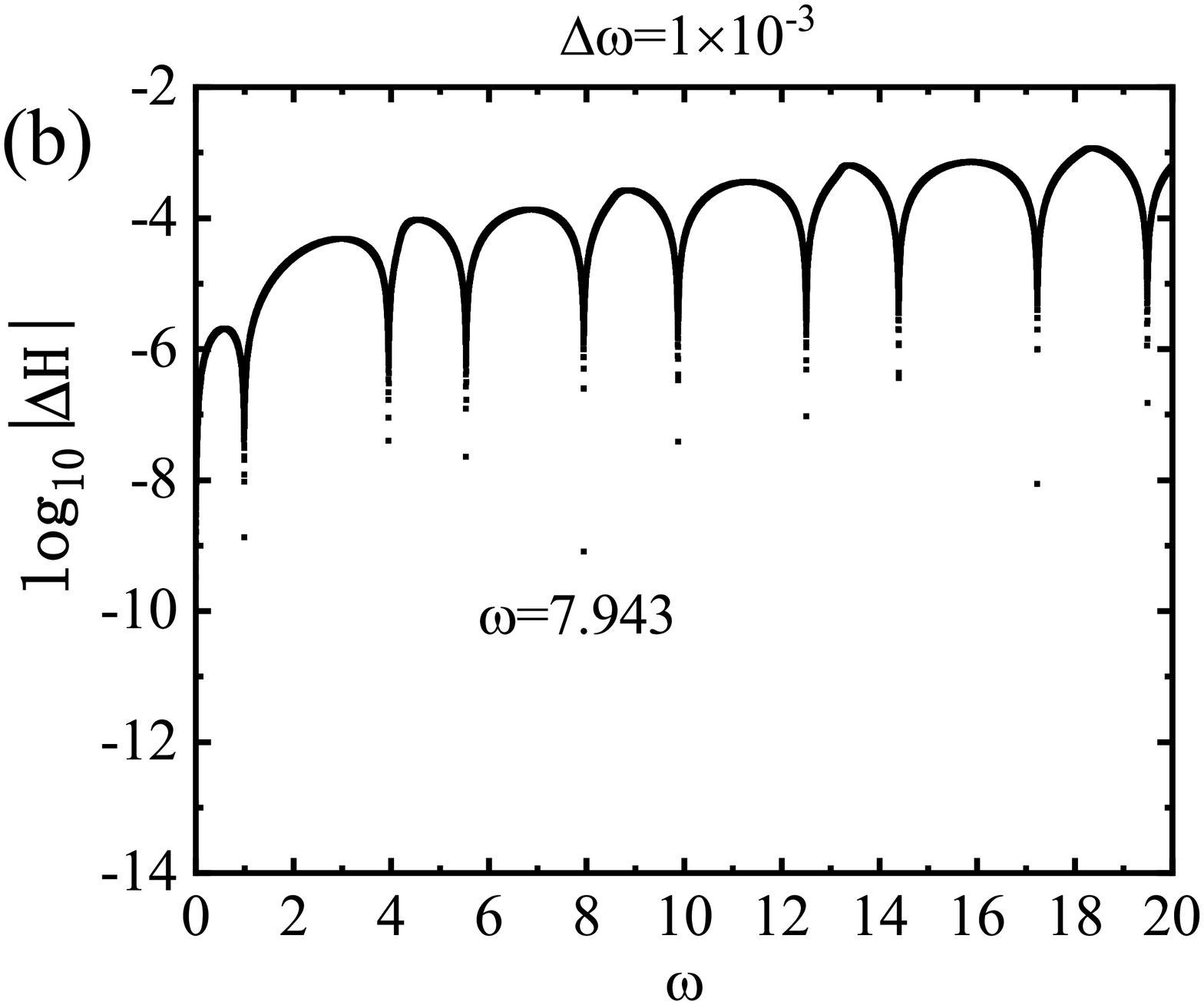}
\caption{Dependence of Hamiltonian error $\Delta H$ on the control
parameter $\omega$. The test orbit is that of Figure 2(a). Let
$\omega$ range from 0 to 20 with an interval $\Delta\omega=0.001$.
Given a value of $\omega$, the error is obtained after $10^7$
integration steps. $\omega=3.553$ corresponds to the best accuracy
for TS4 in panel (a), and $\omega =7.943$ does for TPRK$_{6}$4 in
panel (b).   }}
   \end{figure*}

\end{document}